\documentclass[12pt,reqno]{amsart}

\usepackage{amsmath}
\usepackage{amsfonts}
\usepackage{amsthm}
\usepackage{amstext}
\usepackage{enumerate}

     \newcommand{\supp}{\operatorname{supp}}

     \newcommand{\dist}{{\operatorname{dist}}}

     \newcommand{\Op}{{\operatorname{Op\!^w\!}}}

     \theoremstyle{plain}
     \newtheorem{thm}{Theorem}[section]
     
     \newtheorem{lemma}[thm]{Lemma}
     \newtheorem{cor}[thm]{Corollary}
     \theoremstyle{definition}

     \newtheorem{example}[thm]{Example}
     \newtheorem{assump}[thm]{Assumption}
     
     \newtheorem{remark}[thm]{Remark}
     \newtheorem{remarks}[thm]{Remarks}

     \numberwithin{equation}{section}

\title[Zero energy]{Zero energy asymptotics of the resolvent for a
  class of slowly decaying potentials}
\author{S. Fournais}
\address[S. Fournais]{Laboratoire de Math\'{e}matiques\\
           Universit\'{e} Paris-Sud - B\^{a}t 425\\
           F-91405 Orsay Cedex\\ France.
           }
\email{soeren.fournais@math.u-psud.fr}
\thanks{S. Fournais was supported by a grant from
the Carlsberg Foundation (before
31.12.02) and by a Marie Curie Fellowship of the European 
Community Programme `Improving the Human Research Potential
and the Socio-Economic Knowledge Base' under contract number
HPMF-CT-2002-01822 (from 01.01.03).}

\author{E. Skibsted}
\address[E. Skibsted]{Institut for  Matematiske
Fag \\
Aarhus Universitet\\ Ny Munkegade \\ 8000 Aarhus C \\
Denmark}
\email{skibsted@imf.au.dk}
\thanks{E. Skibsted is (partially) supported by  MaPhySto -- A
Network in Mathematical Physics and Stochastics, funded by
The Danish National Research Foundation.}
\date{\today}

\begin{document}
\begin{abstract} We prove a limiting absorption
principle at zero energy for two-body
Schr\"odinger
operators with  long-range potentials having a positive virial at
infinity. More precisely, we establish a complete asymptotic expansion
of the resolvent in weighted spaces when the spectral parameter varies
in cones; one of the two 
branches of boundary  for the cones being given by the positive 
real axis. The principal tools are absence of eigenvalue at zero, singular Mourre
theory and microlocal estimates.
\end{abstract}

\maketitle

\tableofcontents

\section{Introduction and statement of the main results}
In this paper we derive an asymptotic expansion at zero of
the resolvent $R(\zeta) = (H -\zeta)^{-1}$ of  a two-body
Schr{\"o}dinger operator $H = -\Delta + V$  on $L^2({\mathbb
R}^d)$. It is well-known, see \cite{rauch}, 
\cite{JensenKato} and the more recent work
\cite{JensenNenciu} in which further references can be
found, that if $V(x)=O(|x|^{-(2+\epsilon)})$ (for
$\epsilon>0$) then such an asymptotic expansion exists. The
result is rather complicated; it depends on the dimension
$d$ and on the possible existence of zero-energy eigenstates
and/or a re\-sonance state. For the `long-range' case,
$V(x)=O(|x|^{-\mu})$ with $\mu<2$, the literature is more
sparse. In fact, the only papers on asymptotics of the
resolvent for such potentials, we are aware of, are
\cite{Yafaev} and \cite{Nakamura} (and \cite{BolleGesSch} for
the purely Coulombic case). In \cite{Yafaev} only radially
symmetric potentials are treated, and some of the
assumptions of \cite{Nakamura} (although not requiring
radial symmetry) appear unnecessarily strong. The purpose of the
present  paper is to prove a limiting absorption principle
at zero, and in fact a complete asymptotic expansion of
the resolvent, for a much wider class of potentials. Our
basic assumption is a sign condition at infinity,
 \begin{equation}
   \label{eq:negV}
   V(x) \leq -\epsilon
|x| ^{-\mu};\;|x|> R, 
 \end{equation}
and a similar positive virial condition.

For such potentials we prove complete asymptotic expansions (in weighted spaces)
\begin{equation}
  \label{eq:result}
 R(\lambda+(-)i0)\asymp\sum _{j=0}^\infty R^{+(-)}_j\lambda^j\;{\rm
  for}\;\lambda\to 0^+; 
\end{equation}
here $R^{+}_0\ne R^{-}_0$.
We also show that zero is not an eigenvalue. (This is implicit in
\eqref{eq:result}.)
We notice that there is no explicit dimension-dependence or
fractional/inverse powers in $\lambda$. 

It is well-known that for
`long-range' potentials that are  positive at infinity, zero can indeed be an eigenvalue.
This explains one aspect of  the condition
\eqref{eq:negV}. Probably the best intuitive explanation of the  result
\eqref{eq:result} is given in terms of the WKB-ansatz for stationary
solutions to the Schr{\"o}dinger equation
$-\psi^{\prime\prime}+V\psi=E\psi$ in dimension $d=1$
$$\psi\approx 
C_{+}(E-V)^{-\frac{1}{4}}e^{i\int(E-V)^{\frac{1}{2}}dx}+C_{-}(E-V)^{-\frac{1}{4}}e^{-i\int(E-V)^{\frac{1}{2}}dx}.$$
Under the condition \eqref{eq:negV} there is a trace of scattering
theory even for $E\approx 0$ (but $\geq 0$) in the sense that the oscillatory behaviour
survives in this regime (since $\int(-V)^{\frac{1}{2}}dx
\sim|x|^{1-\frac{\mu}{2}} \to \infty$). Moreover, clearly the ansatz suggests that
zero is not an eigenvalue, and it is also suggestive for which weights one needs in \eqref{eq:result}.  We remark that indeed
\eqref{eq:result} can be proved for $d=1$ by WKB-methods, see   \cite{Yafaev}.
 
Let us state our main results precisely.
Let $0<\theta < \pi$ and define
\begin{equation}
  \label{eq:def_Gamma_theta}
\Gamma_{\theta} = \{ z \in {\mathbb C}\setminus \{0\}
\,\big{ | } \,
|z| \leq 1, \,
\arg z \in (0,\theta) \}.
\end{equation}

For a Hilbert space ${\mathcal H}$ (which in our case mostly will be
$L^2({\mathbb R}^d)$) we denote by ${\mathcal B}(\mathcal H)$ the
space of bounded linear operators on ${\mathcal H}$.
A ${\mathcal B}(\mathcal H)$-valued function $B(\cdot)$ on
$\Gamma_{\theta}$ is said to be uniformly H\"{o}lder continuous in 
$\Gamma_{\theta}$ if there exist $C,\gamma>0$ such that 
$$
\|B(z_1)-B(z_2) \|\leq C|z_1-z_2|^{\gamma}\; {\rm{for\; all}}\; z_1,z_2\in
\Gamma_{\theta}.
$$

A main result of the present paper is the following {\it
limiting absorption principle} at zero energy for a Schr{\"o}dinger
operator $H = -\Delta + V$  on $\mathcal H=L^2({\mathbb R}^d)$. We recall
the notation $R(\zeta) = (H -\zeta)^{-1}$. Notice that it is enough to
impose the conditions (\ref{it:assumption1}) and
(\ref{it:assumption3}) near infinity. We will give arguments to this
effect in Section \ref{sec:lim_abs_unperturbed}.

\begin{thm}[Limiting absorption principle]
\label{thm:lap}
Let $V(x)=V_1(x)+V_2(x)$, $x \in {\mathbb R}^d$, be a
real-valued potential.
Suppose there exists $0<\mu <2$ such
that $V$ satisfies the conditions
{\rm
(\ref{it:assumption1})--(\ref{it:assumption_last})} below.

\begin{enumerate}[\quad\normalfont (1)]
   \item \label{it:assumption1} There exists $\epsilon_1 > 0$ such
     that $V_1(x) \leq -\epsilon_1
\langle x \rangle^{-\mu};\;\langle x \rangle = \sqrt{1 + x^2}$.
   \item \label{it:assumption2}     For all $\alpha\in ({\mathbb N} \cup \{0\})^{d}$
there exists
$C_{\alpha} >0$ such that
$$
\langle x \rangle^{\mu+|\alpha|} |\partial^{\alpha} V_1(x)|
\leq C_{\alpha}.$$
\item \label{it:assumption3} 
     There exists $\epsilon_2 > 0$ such that $-|x|^{-2}
\left(x\cdot \nabla (|x|^2 V_1)\right)
\geq -\epsilon_2 V_1$.
   \item \label{compact} $V_2 (-\Delta +i)^{-1}$ is a compact operator on
     $L^2({\mathbb R}^d)$.
   \item \label{it:assumption4} There exists $\delta,C,R > 0$
such that 
$$
                                |V_2(x)| \leq C |x|^{-1-\mu/2-\delta},
$$ 
for $|x| > R$.
   \item \label{it:assumption_last} $V$ satisfies unique continuation
     at infinity (see Assumption
\ref{assump:unique_continuation} in Section \ref{absence}).
\end{enumerate}

Then for all $s\in (\tfrac{1}{2} + \tfrac{\mu}{4}, \tfrac{1}{2} + \tfrac{\mu}{4} + \delta)$
 and all
$0 < \theta < \pi$ the family of operators $B(\zeta)=\langle x
\rangle^{-s}R(\zeta) 
\langle x \rangle^{-s}$ is uniformly H\"{o}lder continuous in 
$\Gamma_{\theta}$. For  $s>3/2(1+\mu/2)$ (in the case
$\delta>(1+\mu/2$)) the H\"{o}lder exponent
 may be
chosen to be $\gamma=1$, and for $s\leq3/2(1+\mu/2)$ to be any
$\gamma<s(1+\mu/2)^{-1}-1/2$.
In particular there exists $C_{s,\theta}>0$
such that
\begin{equation}
  \label{eq:limitbound}
 \sup_{\zeta \in \Gamma_{\theta}}
\left\|
\langle x
\rangle^{-s} R(\zeta)
\langle x \rangle^{-s}\right\| \leq C_{s,\theta}, 
\end{equation}
and the limits

\begin{align*}
\langle x \rangle^{-s} R(0 + i0) 
\langle x \rangle^{-s}
\equiv \lim_{\zeta \rightarrow 0, \zeta \in \Gamma_{\theta}}
\langle x \rangle^{-s} R(\zeta)
\langle x \rangle^{-s},
\end{align*}
\begin{align*}
\langle x \rangle^{-s} R(0 -i0) \
\langle x \rangle^{-s}
\equiv \lim_{\zeta \rightarrow 0, \zeta \in \Gamma_{\theta}}
\langle x \rangle^{-s}
R\left(\bar{\zeta}\right)
\langle x \rangle^{-s}
\end{align*}
exist in
${\mathcal B}(L^2({\mathbb R}^d))$.
\end{thm}

From the facts that $V$ is negative at infinity and decays
slower than $r^{-2}$ it follows, cf. \cite[Theorem XIII.6
(Vol. IV, p. 87)]{ReSi}, that $H$ has infinitely many
negative eigenvalues accumulating at zero. Clearly by
\eqref{eq:limitbound}, zero is not an eigenvalue of $H$. (See Section
\ref{absence} for a more general result.)

We also get existence of limits for powers of the resolvent.
The a\-symptotic expansion \eqref{eq:result} is an easy
consequence of Theorem \ref{thm:lap_iterated} below.

\begin{thm}[Iterated resolvents]
\label{thm:lap_iterated}
Let $V = V_1 + V_2$ satisfy the conditions in Theorem
\ref{thm:lap} with {\rm (5)} replaced by
\begin{itemize}
\item[(5')] $\supp (V_2)$ is compact.
\end{itemize}
Let $H = -\Delta + V$ and $R(\zeta) = (H -\zeta)^{-1}$ be given as before
and 
$k(x) =
\langle x
\rangle^{1+\mu/2}$, $x \in {\mathbb R}^d$. Let $m
\in {\mathbb N}$, $\theta \in (0,\pi)$ and $\epsilon
>0$. Then there exists $C > 0$ such that
\begin{equation}
  \label{eq:iterated_estimate}
  \left\| k^{-(m-1/2) -\epsilon} R(\zeta)^m k^{-(m-1/2) -\epsilon}
  \right\| \leq C,
\end{equation}
for all $\zeta \in \Gamma_{\theta}$. \\
Furthermore, the function
$$
\zeta \mapsto k^{-(m-1/2) -\epsilon} R(\zeta)^m k^{-(m-1/2)
-\epsilon},
$$
extends to a continuous function on
$\overline{\Gamma}_{\theta}$ (the closure of $\Gamma_{\theta}$).
\end{thm}

\begin{remarks}
\begin{enumerate}[\normalfont 1)]
\item  For fixed $m \in {\mathbb N}$ the bound
$V_2=O(k^{-m-\epsilon})$ suffices for
\eqref{eq:iterated_estimate}. (This follows readily from
our proof by  a little more bookkeeping.) 
\item In dimension $d=1$ one may show by WKB-analysis, see
for example \cite{Yafaev}, that \eqref{eq:iterated_estimate}
is optimal in the following sense: There exists  $\phi \in
L^{d}({\mathbb
  R}^d)$  such that  
\begin{equation*}
  \label{eq:1yy}
  \sup_{\zeta\in \Gamma_{\theta}}\left\| k^{-(m-1/2)} R(\zeta)^m \phi
  \right\| =\infty.
\end{equation*}
(This may be done using $
R(\zeta)^m = ((m-1)!)^{-1} \frac {d^{m-1}}{d\zeta^{m-1}} R(\zeta)$ and the  standard formula for $R(\zeta)$ in terms of outgoing/incoming
ge\-neralized eigenfunctions; analysing  the large $x$-asymptotics of
the $(m-1)$'th derivative w.r.t. $\zeta$ of those functions yields the
result.) We would expect that  the same result is true in any dimension, also for potentials that are not radially
symmetric.

\end{enumerate}
\end{remarks}

Due to Theorem \ref{thm:lap} we may define
\begin{align*}
 E^{\prime}(+0)= &(2\pi i)^{-1}\left( R(0+i0)-R(0-i0)\right)
\in {\mathcal B}({\mathcal H}_1, {\mathcal H}_2);\\&
{\mathcal H}_1=k^{-1/2 -\epsilon}L^2({\mathbb R}^d),\;{\mathcal
   H}_2=k^{1/2 +\epsilon}L^2({\mathbb R}^d).
\end{align*}
Motivated by the next result we  shall prove (in Section
\ref{iterated_perturbed}) that 
\begin{eqnarray}
  \label{eq:E ne 0}
  E^{\prime}(+0)\ne 0.
\end{eqnarray}
Let $F(|x|<C)$ denote the multiplication operator by the
characteristic function of $\{x|\;|x|<C\}$.

\begin{thm}[Local decay  estimates]
\label{thm:minimal velocity} Under the conditions of Theorem \ref{thm:lap_iterated}:
\begin{enumerate}[\normalfont (i)]
\item\label{it:local decay estimate} For all $s>\frac {5}{2}(1+\frac
  {\mu}{2})$
and $f\in C^{\infty}_0(\mathbb R)$
$$\|\langle x
\rangle^{-s}\Big(e^{-itH}(f1_{[0,\infty)})(H)+it^{-1}f(0)E^{\prime}(+0)\Big)\langle x
\rangle^{-s}\|=O(t^{-2}).
$$
\item\label{it:minimal velocity bound}
  For all  $0\leq\epsilon^{\prime}<\epsilon\leq 1$, all real $s$ such
  that for some integer $m$ 
$$s(1+\mu/2)^{-1}+\frac {1}{2}>m>\frac {1}{2}+{\epsilon}^{-1}(1+\frac {1}{2}\epsilon^{\prime}),$$
 and for all $f\in C^{\infty}_0(\mathbb R)$
\begin{align}
  \label{eq:minimal veocity bound}
  &\| F(|x|<t^\kappa)e^{-itH}(f1_{[0,\infty)})(H)\langle x
\rangle^{-s}\|=O(t^{-(1+\epsilon^\prime)\frac {1}{2}});\\
&\kappa={(1-\epsilon)(1+\mu /2)^{-1}}.\nonumber
\end{align}
\end{enumerate}
\end{thm}

\begin{remarks}
\begin{enumerate}[\normalfont 1)]
\item By time reversal invariance there are similar bounds for
  $t\to -\infty$.
\item Due to \eqref{eq:E ne 0} and Theorem \ref{thm:minimal
velocity} (\ref{it:local decay estimate}) the best one could
hope for to the right in \eqref{eq:minimal veocity bound}
would be the bound $O(t^{-1})$ (for $f(0)\ne 0$). Moreover we
would expect that $\kappa=(1+\mu /2)^{-1}$ is indeed the
borderline for this kind of low energy, minimal velocity
estimate; see Theorem \ref{thm:classicalminimal velocity
bound} for an analogous bound in classical mechanics.
\item If $V_2\in C^\infty_0({\mathbb R}^d)$ one may take
$f=1$ in Theorem \ref{thm:minimal velocity} (\ref{it:local
decay estimate}) and (\ref{it:minimal velocity bound}). This follows
 readily from  the given statements and  well-established high energy
estimates, see \cite [Theorem 1.1]{Kitada}, \cite [Theorem 1]{cycper} or 
\cite [Theorem 1.2 (ii)]{Jensen2}. Some local singularities
may be included (see for example \cite [Section 6]{Jensen2} for a
specific treatment of the purely Coulombic case of the Hydrogen
atom).
\end{enumerate} 
\end{remarks}

There are other (main) results that are in fact important in the proof
of the above Theorems \ref{thm:lap} and \ref{thm:lap_iterated}. They
are however too complicated to be stated in this introduction. They
concern {\it {microlocal estimates}} of the resolvent. 

One perspective of our results is  zero-energy asymptotics of the
scattering matrix (or related quantities) for the class of potentials consi\-dered in this
paper.
It is well-known that  microlocal estimates are important in the
study of the 
scattering matrix for positive energies (and in the high energy
regime), cf. the seminal work  \cite
{IsozakiKitada}. We would hope that our estimates would open up for 
results on asymptotics of scattering quantities; this needs
elaboration elsewhere.

Let us also mention the perspective of asymptotics at thresholds for
the many-body problem. This appears, although very interesting, far
more ambitious. In the time-dependent approach to the asymptotic
completeness problem control of the dynamical behaviour at
thres\-holds is important. We were  motivated to look at the above
two-body problems by the
papers \cite{Ge} and \cite{Skib} on three-body asymptotic
completeness with long-range potentials; some estimates, similar in
nature to \eqref{eq:minimal veocity bound}, play an important role in
these papers. 

To minimize confusion, let us mention a number of
(standard) notations and conventions that will be used in the
paper. 

We will write $\Re z  = \Re(z), \Im z = \Im(z)$ for the real and imaginary
part of $z$, both in the case where
$z$ is a complex number and, more generally, when it is an
operator.

We define for any open $U \subseteq {\mathbb R}^d$
$$
{\mathcal B}^{\infty}(U) = 
\left\{ f\in C^{\infty}(U) \, \Big{|} \, \partial^{\alpha} f \in
  L^{\infty}(U) \text{ for all } \alpha \in ({\mathbb N} \cup \{0\})^{d} 
\right\}.
$$

Apart from the notation 
$\langle x \rangle = \sqrt{1 + x^2}$, used above, we will also need
the notation $p = -i \nabla$ and $A = (x\cdot
p + p \cdot x)/2$.

The virial $W$ of the potential $V$ will play an important role 
throughout the paper. It is defined by
\begin{equation}
  \label{eq:definition_virial}
  W = -2V - x\cdot \nabla V.
\end{equation}
We recall the (formal) identity  $i[H,A]=2H+W$.
The assumptions (\ref{it:assumption1}) and
(\ref{it:assumption3}) of Theorem \ref{thm:lap} in the case $V_2=0$
yield
\begin{equation}
  \label{eq:definition_virial2}
W(x)
\geq \epsilon_1 \epsilon_2 \langle x \rangle^{-\mu},  
\end{equation} 
so in particular, $W$
is positive in this case.

In Section \ref{absence} we will prove a result of independent
interest (Theorem \ref{thm:absence}), namely the
non-existence of eigenvalue at zero under slightly weaker
assumptions than those of Theorem
\ref{thm:lap}. This is a basic result which will allow us
to prove Theorem
\ref{thm:lap}  by a perturbative method: First we shall prove
\eqref{eq:limitbound} in the case
$V_2 =0$, and then later on in ge\-neral by perturbation theory and Theorem
\ref{thm:absence}. We prove Theorem \ref{thm:absence} by extending
the method of proof of 
\cite[Theorem XIII.58]{ReSi} dealing with absence of positive
eigenvalues to incorporate the case $E=0$.

The bound
\eqref{eq:limitbound} of Theorem
\ref{thm:lap} in the case $V_2 =0$ is shown in Section
\ref{sec:lim_abs_unperturbed}---this is done using a non-standard
Mourre theory.

In Section \ref{iterated} we
prove Theorem \ref{thm:lap_iterated} in the case $V_2 =0$.
The proof is
accomplished using the strategy from \cite{GIS} and an
energy-dependent pseudodifferential calculus. Some technical
verifications concerning this calculus
will be given in Appendix \ref{verification}. In Appendix
\ref{algebra}  a certain `algebraic verification' is given.
Using Theorems \ref{thm:lap_iterated} (with $V_2 =0$) and 
\ref{thm:absence} we finally prove Theorem \ref{thm:lap} in the
general case in Section
\ref{iterated_perturbed}. That section also contains the proof of
Theorem \ref{thm:lap_iterated}, \eqref{eq:E ne 0} and Theorem \ref{thm:minimal velocity}. 

\section{Absence of eigenvalue at zero}
\label{absence}
In this section we will prove a basic result,
namely that for long-range potentials $V$ that are  {\it negative} at
infinity, there is no $L^2$-eigenfunction with energy zero. That
is the result of Theorem \ref{thm:absence}. In Subsection \ref{snik-snak}
we will use this insight to obtain the
limiting absorption principle at zero energy, Theorem
\ref{thm:lap}.

We
intend to generalize the `Kato-Agmon-Simon Theorem'
\cite[Theorem XIII.58]{ReSi} by using a modification of its
proof. In particular, we shall apply unique continuation. The proof
uses ODE techniques in the radial coordinate, so let us specify the
notation $x = r \omega \in {\mathbb R}^d$, with $\omega \in {\mathbb
S}^{d-1}$.

The conditions which exclude zero-energy
eigenfunctions are given in Assumptions
\ref{assump:unique_continuation} and  \ref{assump:Erik_conditon_1.2}.

\begin{assump}[Unique continuation at infinity]
\label{assump:unique_continuation}
The function $V: {\mathbb R}^d \rightarrow {\mathbb R}$ is measurable,
and  if $u \in
H^2({\mathbb R}^d)$, $u =0$ in a neighbourhood of $\infty$, the product $V \psi \in
     L^2({\mathbb R}^d)$ and $u$ is a distributional solution to
$$
-\Delta u + V u = 0,
$$
then $u=0$.
\end{assump}

\begin{remark}
  Different results exist in this direction. The condition $V \in L^{d/2}_{\rm{loc}}({\mathbb
  R}^d)$ suffices for $d\geq 3$, see \cite{JeKe}.
\end{remark}

\begin{assump}
\label{assump:Erik_conditon_1.2}
The function $V$ can be written as $V= V_1 + V_2$, such that:
For some $s \in [0,1)$, some $R,C > 0$ and a positive
differentiable function $h = h(r)$ defined on $[R,\infty)$ we
have
\begin{enumerate}[\quad\normalfont (1)]
   \item $V_1$ and $V_2$ are bounded on $|x| > R$, and $V_1$
is negative
     on $|x| > R$.
   \item $\sup_{\omega \in S^{d-1}} \frac{d}{dr} (r^{s+1} V_1(r\omega))
     \leq -r^s h^2(r)$ when $r > R$.
   \item \label{item3} $r^{-1} + r\sup_{\omega \in S^{d-1}} |V_2(r\omega)| = o(h)$ as
     $r \rightarrow \infty$.
   \item $h'(r) \leq C h^2(r)$ on $|x| > R$.
\end{enumerate}
\end{assump}

With the above  assumptions we can prove the absence of
zero-energy eigenstates.

\begin{thm}
\label{thm:absence}
Suppose $V = V_1 + V_2$ satisfies
Assumptions
\ref{assump:unique_continuation}
and \ref{assump:Erik_conditon_1.2}. Suppose furthermore that
$\psi \in H^2_{\rm{loc}}({\mathbb R}^d)$ satisfies {\rm
(\ref{it:assumption111})--(\ref{it:assumption_last111})} below.
\begin{enumerate}[\quad\normalfont (1)]
   \item \label{it:assumption111} $\int_{|x| > R} h^2(r) |\psi(x)|^2 \,dx  < \infty$
and $\int_{|x| > R}|V_1(x)| |\psi(x)|^2 \,dx  < \infty$.
   \item \label{it:assumption112} $p_j \psi  \in L^2({\mathbb
       R}^d);\;j=1,\cdots ,d.$
   \item \label{it:assumption_last111} The product $V \psi \in
     L^2_{\rm{loc}}({\mathbb R}^d)$, and 
   $(-\Delta + V) \psi  = 0$ in the sense of distributions.
\end{enumerate}

Then $\psi = 0$.
\end{thm}

\begin{remarks} \label{h}
\begin{enumerate}[\normalfont 1)]
\item \label{it:hepsilon}The assumptions in Theorem \ref{thm:lap} are
stronger than Assumption \ref{assump:Erik_conditon_1.2}: Take  $h =
\epsilon r^{-\mu/2}$ 
for a small $\epsilon > 0$ and 
$s$ close to $1$.
\item \label{it:Rresult} Suppose Assumption  \ref{assump:Erik_conditon_1.2} and
  that $\psi \in H^2_{\rm{loc}}({\mathbb R}^d)$ obeys
  (\ref{it:assumption111}), (\ref{it:assumption112}) and
  (\ref{it:assumption_last111}) except that  the condition $(-\Delta + V) \psi  = 0$
  is only required to be fulfilled outside some sufficiently large ball $B_\rho=\{x\in{\mathbb
    R}^d|\;|x|\leq \rho
\}$, then our proof yields the conclusion that $\psi = 0$ outside
$B_\rho$. (This remark will 
be used in Subsection \ref{snakk}.)
\item Using the negativity of $V_1$ we see that
   \begin{align}
       \label{eq:auxilliary}
       \sup_{\omega \in S^{d-1}} \frac{d}{dr} r^2 V_1(r\omega)
       \leq -r
       h^2(r); \text{ for } r > R.
   \end{align}
\item By integrating \eqref{eq:auxilliary} we get the bound 
$$
\limsup_{\rho\rightarrow \infty} \rho^{-2} \int_R^{\rho} rh^2(r)dr \leq
\sup_{|x|>R} |V_1|,
$$
which essentially is a boundedness condition on $h$.
\item A slightly more general result may be obtained in
terms of an $h$ that depends on the angle $\omega$ as well.
\end{enumerate}
\end{remarks}

\begin{example}
Suppose that for some $\mu \in [0,2)$, $V_1(x)
= v(\omega)r^{-\mu} + R(x)$  where $\sup v(\omega) <0$,
$R(x) = o(r^{-\mu})$ and  $\sup_{\omega \in S^{d-1}}
\frac{d}{dr}R(r\omega) = o(r^{-\mu -1})$, and
$V_2 = o(r^{-1-\mu/2})$.
Then Assumption \ref{assump:Erik_conditon_1.2} holds, cf. Remarks
\ref{h} \ref{it:hepsilon}.
The particular case given by putting
$\mu =0$ and $v$ equal to a negative constant (in that case
we may take $s=0$ in Assumption \ref{assump:Erik_conditon_1.2}) yields the Kato-Agmon-Simon theorem on absence
of positive eigenvalues for Schr\"odinger operators \cite[Theorem XIII.58]{ReSi}. (Notice
that any positive energy may be
absorbed into $V_1$.)
\end{example}

\begin{proof}[Proof of Theorem \ref{thm:absence}]
The proof is based on unique continuation, so we aim at proving that
$\psi$ vanishes identically on ${\mathbb R}^d \setminus \{|x| \leq R_1 \}$
for some (sufficiently big) $R_1$.
Let $B$ be the (negative) Laplace-Beltrami operator on
${\mathbb S}^{d-1}$ such that
$$
-\Delta = -\frac{\partial^2}{\partial r^2}
-\frac{d-1}{r}\frac{\partial}{\partial r}
-\frac{1}{r^2} B.
$$
We will need the inner product $\langle \cdot, \cdot
\rangle_{L^2({\mathbb S}^{d-1})}$ and the associated norm
$\| \cdot \|_{L^2({\mathbb S}^{d-1})}$ repeatedly. For
shortness we will therefore just write $\langle \cdot, \cdot
\rangle$ and $\|\cdot \|$. Unless otherwise stated, all
inner products and norms in the proof will refer to
these---this is contrary to the convention in the later sections
of the paper where inner products and norms are on
$L^2({\mathbb R}^{d})$ unless otherwise stated.

Let
$\psi$ be the function from the statement of the theorem. We may
assume that $\psi$ is real-valued. We
define, for
$r \in (0,\infty)$, the function
$w_{\psi}(r,\cdot) = w(r,\cdot) \in L^2({\mathbb S}^{d-1})$ by
$$
w(r,\omega) = r^{(d-1)/2} \psi(r\omega).
$$
Notice, that if $\phi \in C_0^{\infty}({\mathbb R}^d)$, then
\begin{equation}
\label{eq:H_forneden}
w_{H\phi} = - w_{\phi}'' - r^{-2}B w_{\phi} +
\tfrac{1}{4}(d-1)(d-3)r^{-2}w_{\phi} + V w_{\phi}.
\end{equation}
  We define furthermore,
   \begin{align}
     \label{eq:def_of_F}
     F(r) = \|w'\|^2 + r^{-2} \langle w, B w \rangle - \langle w, V_1
     w\rangle - sr^{-1} \langle w, w'\rangle,
   \end{align}
where $w'(r,\omega) = \frac{\partial}{\partial
r}w(r,\omega).$

One may note that since $V$ is locally bounded outside a
sufficiently large ball centered at  the origin, one gets by
elliptic regularity (see for instance
\cite{Gilbarg&Trudinger}) that
$\psi \in C^{1,\alpha}_{{\rm loc}}({\mathbb R}^d\setminus
\{ |x| \leq R\})$ (for $R$ given as in Assumption \ref
{assump:Erik_conditon_1.2} and for all $\alpha
\in [0,1)$). Thus
$F$ is a H\"older continuous function.

The first important step in the proof is to establish that 
$F$ is integrable at infinity, i.e. $F \in L^1((R,\infty);dr)$.
In order to see this, notice that for $\phi \in
C_0^{\infty}({\mathbb R}^{d})$ we have
$$
\int_0^{\infty} \left\{ \| r^{(d-1)/2}
\frac{\partial}{\partial r} \phi (r\omega)\|^2 - r^{-2}
\langle w_{\phi}, B w_{\phi}
\rangle \right\} \,dr= \| |p|\phi \|^2_{L^2({\mathbb
R}^d)}.
$$
This implies, since $-B \geq 0$,
\begin{align*}
\int_R^{\infty} \| w_{\phi}' \|^2 \, dr &
\leq C \int_R^{\infty} \left(
\| r^{(d-1)/2} \frac{\partial}{\partial r}\phi(r\omega) \|^2
+r^{-2}\| w_{\phi} \|^2 \right) \,dr \\
& \leq C' \left( \| |p|\phi \|^2_{L^2({\mathbb R}^d)}
+
\int_{|x| \geq R} h(r)^2 |\phi(x)|^2 \,dx \right).
\end{align*}

This, and an approximation argument, proves that the first term in \eqref{eq:def_of_F} is
integrable. Similar considerations apply to the other terms. Thus
$F\in L^1((R,\infty);dr)$. 

We will next compute $(rF)'$ using that formally
\begin{align}
\label{eq:ligning_w}
0 = w_{H\psi} = - w'' - r^{-2}Bw + \tfrac{1}{4}
(d-1)(d-3)r^{-2}w + V w.
\end{align}
Using \eqref{eq:ligning_w}, a formal computation gives
\begin{align}
   \label{eq:formal_1}
   (r F(r))' & =
   2 \langle w', [rV_2 + \tfrac{1}{4}(d-1)(d-3)r^{-1}]w
\rangle + (1-s) \|w'\|^2 \nonumber \\
&-\langle w, (rV_1)' w
\rangle - (1-s) r^{-2} \langle w, B w\rangle \nonumber \\
&-
s \langle w, [\tfrac{1}{4}(d-1)(d-3)r^{-2}+V] w \rangle.
\end{align}
Let
$\delta = 1-s >0$. Using Cauchy-Schwarz, we estimate the
first term in
\eqref{eq:formal_1} below by
\begin{align}
  \label{eq:first_below}
  - \delta \| w'\|^2 - \langle w, o(h^2) w \rangle.
\end{align}
For the third term we estimate
\begin{align}
\label{eq:second_below}
   -(rV_1)' = -r^{-s} \frac{d}{dr}(r^{s+1} V_1) + sV_1 \geq
h^2 + sV_1.
\end{align}
 Notice that the contribution from $sV_1$ to the right and the
 contribution from the term $V_1$ in the fifth term of
 \eqref{eq:formal_1} in fact cancel. Clearly the fourth term is
 non-negative. Hence we conclude from \eqref{eq:formal_1}, \eqref{eq:first_below}
and \eqref{eq:second_below} that for a sufficiently large 
$R_1\geq R$
\begin{align}
   \label{eq:ERIK_2.2}
  \frac{d}{dr}(rF(r)) & \geq \langle w, (h^2 + o(h^2)) w
\rangle  \geq 0 \text{ when } r> R_1.
\end{align}
Therefore we have, for all $r>r_1>R_1$,
\begin{equation}
\label{eq:stable_conclusion}
F(r) \geq \frac{r_1F(r_1)}{r}.
\end{equation}

The deduction of \eqref{eq:stable_conclusion} above was
based on formal calculations. However, it is standard to
use approximation arguments (see the proof of \cite[Thm.
XIII.58]{ReSi}) to justify the inequality rigorously. Below
we will do a similar formal calculation without
comment---again it can easily be justified using
approximations by smooth functions.

Since $F \in L^1(dr)$ it follows from
\eqref{eq:stable_conclusion} that
\begin{align}
   \label{eq:negativity_of_F}
   F(r) \leq 0 \text{ for all } r > R_1.
\end{align}

Next we define, with $w_m = r^m w$ for $m > 0,$
\begin{multline}
     \label{eq:def_of_G}
     G(m,r) = \|w_m'\|^2 + r^{-2} \langle w_{m}, B w_{m}
    \rangle \\
    +
     \langle w_{m}, (m(m+1)r^{-2}-g-V_1) w_{m}\rangle,
\end{multline}
where $g = \epsilon r^{-1}h(r)$ with $\epsilon = (2C)^{-1}$.
Here $C$ is the constant from Assumption
\ref{assump:Erik_conditon_1.2}.
The function $w_m$ satisfies
\begin{align*}
&w_m'' -2mr^{-1} w_m' + r^{-2}\left(m(m+1) -
\tfrac{1}{4}(d-1)(d-3)+B\right)w_m - V w_m  \\
&= 0.
\end{align*}
We now compute,
using~\eqref{eq:auxilliary} in the first inequality, for
$r>\tilde{R}_1$ and $m > M$, for some large positive
$\tilde{R}_1$ and $M$:
\begin{align}
   \label{eq:Erik_2.5}
&(2r)^{-1} \frac{d}{dr}(r^2 G(m,r)) \nonumber \\
& =
\langle w_m, (rV_2-rg + \tfrac{1}{4}(d-1)(d-3)r^{-1}) w'_{m}
\rangle
\nonumber \\
&\quad  + (2m+1) \|w'_m \|^2 - \langle w_m, (2r)^{-1} (r^2
(g+V_1))'w_m
\rangle \nonumber \\
& \geq (2m+1) \|w'_m \|^2
+ (\tfrac{1}{2}h^2-(2r)^{-1}(r^2g)') \|w_m
\|^2 \nonumber \\
&\quad + \langle w_m, (o(h)-rg) w'_m \rangle \nonumber \\
& = (2m+1) \|w'_m \|^2 + (\tfrac{1}{4}h^2 + o(h^2)) \|w_m
\|^2 + \langle w_m, O(h) w'_m \rangle \nonumber \\
& \geq 0; \text{ for } r>\tilde{R}_1, m > M.
\end{align}
For convenience we assume henceforth that $\tilde{R}_1 =
R_1$. We learn from \eqref{eq:Erik_2.5} that $r^2 G(m,r)$ is
non-decreasing in $r > R_1$, provided that $m > M$.

We will now combine \eqref{eq:negativity_of_F} and
\eqref{eq:Erik_2.5} to prove that $w$ (and therefore
$\psi$) vanishes in a neighbourhood of infinity. Clearly
\begin{align}
\label{eq:Erik_2.6}
   & r^{-2m} G(m,r) =  \\
   & \| w'+mr^{-1}w \|^2 + r^{-2} \langle w, B w \rangle + \langle w,
   (m(m+1)r^{-2}-g-V_1)w \rangle.\nonumber
\end{align}
Suppose $w(r_1) \neq 0$ for some $r_1 > R_1$. Then we
have for some large enough $m_1 > M$,
\begin{align}
   \label{eq:Erik_2.7}
   r_1^{-2} \langle w(r_1), B w(r_1) \rangle + \langle w(r_1),
   (m_1(m_1+1)r_1^{-2}-g(r_1))w(r_1) \rangle > 0.
\end{align}
Clearly \eqref{eq:Erik_2.5}, \eqref{eq:Erik_2.6} and 
\eqref{eq:Erik_2.7} yield
\begin{align}
\label{eq:Erik_2.8}
   G(m_1,r) > 0 \text{ for all } r \geq r_1.
\end{align}
Next, we get from Assumption
\ref{assump:Erik_conditon_1.2} (\ref{item3}) that 
$\lim_{r \rightarrow \infty} r^2 g(r) = \infty$.
Therefore, for some $R_2 > r_1$, we have
\begin{align}
\label{eq:Erik_2.9}
   \left[ \frac{2m_1 + s}{2} + m_1(2m_1+1)\right] r^{-2} -
g(r)
\leq 0
\text{ for all } r > R_2.
\end{align}
Since by assumption 
$$
\int_{r \geq R_2} r^{-
2} \| w(r) \|^2\,dr <  \infty,
$$
 the function $r \mapsto r^{-1} \| w(r)
\|^2$ cannot be strictly increasing near infinity. We pick
$r_2 > R_2$, such that
$\left(r^{-1} \| w(r) \|^2\right)' \big{|}_{r=r_2} \leq 0$ and
estimate using 
\eqref{eq:Erik_2.8} and  \eqref{eq:Erik_2.9},
\begin{align*}
   0 < r_2^{-2m_1} G(m_1,r_2) \leq F(r_2),
\end{align*}
which by \eqref{eq:negativity_of_F} is impossible.

We conclude that $w(r) = 0$ for all $r > R_1$.
Since therefore $\psi = 0$ on a neighbourhood of infinity,
$\psi \in H^2({\mathbb R}^d)$ and $V\psi \in L^2({\mathbb R}^d)$. From Assumption
\ref{assump:unique_continuation} we conclude that $\psi=0$. 
\end{proof}

\section{Limiting absorption principle for $V_1$}
\label{sec:lim_abs_unperturbed}

In this section we will prove the limiting absorption
principle bound \eqref{eq:limitbound} in the case  $V_2=0$ in the set
of conditions of Theorem \ref{thm:lap}. A somewhat more general  result will
be stated in Corollary \ref{cor:unperturbed} below. Explicitly,
we study the operator
$H_1= p^2 + V_1$, where $V=V_1$ satisfies
the assumptions of Theorem \ref{thm:lap}. 
In Section
\ref{iterated_perturbed} we will make a perturbative argument to
include $V_2$ (the short-range, non sign-definite part of
the potential). For simplicity we will skip the indices in
this section and write $H$  and $V$ instead of $H_1$  and
$V_1$, respectively. Our approach is very different from
\cite{Nakamura}, where a similar (but much weaker) result is
proved. There an auxiliary operator (the inverse of the
Birman-Schwinger kernel) is studied, whereas we do Mourre
theory directly on $H$. Since we do not have a positive
commutator in the sense of Mourre \cite{mourre} at
zero-energy, this will be a non-standard Mourre theory. (For another 
non-standard Mourre theory, used in a very different setting, we refer to 
\cite{Herbst91}.)

To motivate our strategy of proof of Theorem \ref{thm:lap} we notice that if $V=V_1+V_2$ only satisfies
the bounds of the assumptions (\ref{it:assumption1}) and (\ref{it:assumption3}) for
$|x| > R$ (and in addition (\ref{it:assumption2}), (\ref{compact}) and
(\ref{it:assumption4})) then there exists another decomposition $V =
\tilde V_1+ \tilde V_2$ for which
(\ref{it:assumption1})--(\ref{it:assumption4}) hold.
To prove this we define $\tilde V_1 = V_1 \chi_{+}(r/n) -  \epsilon_1\langle x \rangle^{-\mu}
\chi_{-}(r/n)$. Here $1=\chi_{+}+ \chi_{-}$ is a standard, smooth
partition of unity on ${\mathbb R}_{+}$: $\supp \chi_{+} =
[1,\infty)$, $\chi_{+}(t) = 1$ for $t \geq 2$. Clearly we may
assume that $\epsilon_2<2-\mu$.
We claim  that for any $n>R$
\begin{align}
  \label{eq:W}
  -x\cdot \nabla \tilde V_1 & \geq (2-\epsilon_2) \tilde V_1.
\end{align}
Using that \eqref{eq:W} is satisfied for $V_1$ on $|x| > R$, we find
\begin{align}
  \label{eq:x_nablaW}
  -x\cdot \nabla \tilde V_1 & \geq (2-\epsilon_2) \tilde V_1
  \nonumber \\
    &
  +\epsilon_1\langle x \rangle^{-\mu} \chi_{-}(r/n)\left( (2-\epsilon_2)
  +
  \langle x \rangle^{\mu}(x\cdot \nabla \langle x
  \rangle^{-\mu})\right)  \nonumber \\
  &-\frac{r}{n} \chi_{-}'(r/n)\langle x \rangle^{-\mu}\left( - \langle
  x \rangle^{\mu} V_1 - \epsilon_1 \right)\nonumber \\
& \geq (2-\epsilon_2) \tilde V_1,\nonumber 
\end{align}
proving \eqref{eq:W}.

We define the operator $\langle A \rangle = (C +
A^2)^{1/2}$ with $C \gg 1$ fixed (yielding better
commutation properties than with $C=1$, cf. \cite[Lemma 2.5]{MoSk}). 
For $E \geq 0$, let $f_{E}$ be the function
\begin{equation}
  \label{eq:def_f}
  f_{E} = \left(E + \langle x \rangle^{-\mu}\right)^{1/2}.
\end{equation}
For the purpose of proving \eqref{eq:limitbound} one could
replace $f_E$ by $\langle x \rangle^{-\mu/2}$ e\-verywhere in
this section. But we include it here since the estimates we
establish (with $f_E$) will be important in Section
\ref{iterated}. 

It is easy to see that $p^2 + f_{E}^2$ defines a positive,
self-adjoint operator, so we can use the spectral theorem to define,
for $\zeta \in {\mathbb C}$, the operator
\begin{align}
  \label{eq:def_gamma}
\gamma = \gamma_{|\zeta|} = (p^2 + f_{|\zeta|}^2)^{1/2},
\end{align}

Recall from \eqref{eq:definition_virial} and
\eqref{eq:definition_virial2} that
$$
W = -2V - x\cdot \nabla V\geq c \langle x \rangle^{-\mu}.
$$
Consider
\begin{equation}
  \label{eq:def_R_zeta}
  R_{\zeta}(\epsilon) = (H - i\epsilon i[H, A] - \zeta)^{-1},
\end{equation}
in a  range of the form 
\begin{equation}
\label{eq:conditions_on_epsilon_and_z}
0<\Im \zeta, \;|\zeta|\leq 1,\;
-2\epsilon_0 \Re \zeta \leq \Im \zeta;\;0<\epsilon\leq
\epsilon_0^\prime\leq \epsilon_0.
\end{equation}
(The existence of the inverse in the definition of
$R_{\zeta}(\epsilon)$ follows from the calculation \eqref{eq:signn}
and the theory of
numerical range \cite[Section V.3.1]{Kato}.) The results
given below have completely analogous versions with the signs of $\Im \zeta$ and
$\epsilon$ both negative; for convenience we shall only
explicitly state the results for  the case of positive signs.

We first formulate and prove a  version of the quadratic
estimate of \cite{mourre}.

\begin{lemma}
  \label{lem:quadratic}
For all $\epsilon_0>0$ there exists $\epsilon_0^\prime>0$ such that
for all positive 
$\epsilon\leq \epsilon_0^\prime$ and all $\zeta$ as in
\eqref{eq:conditions_on_epsilon_and_z}, we have with $\gamma =
\gamma_{|\zeta|}$ given by
\eqref{eq:def_gamma} and for any bounded operator $B$
\begin{align}
\label{eq:quadratic}
\|\gamma R_{\zeta}(\epsilon) B \|^2 \leq C \epsilon^{-1} \|B^*
R_{\zeta}(\epsilon) B \|.
\end{align}
\end{lemma}

\begin{remark}
Note that $\zeta$ appears twice in \eqref{eq:quadratic}: In
the definition of $\gamma = \gamma_{|\zeta|}$ and in
$R_{\zeta}(\epsilon)$.
\end{remark}

\begin{proof}
  Let $T = \gamma R_{\zeta}(\epsilon) B$. Then by the definition
  of $\gamma$
$$
T^* T = (R_{\zeta}(\epsilon) B)^* (p^2 + |\zeta| + \langle x
\rangle^{-\mu}) R_{\zeta}(\epsilon) B.
$$
Notice that with $C_1 = 2C_2 -1$ we have, using $i[H,A]= 2H + W$,
\begin{align}
\label{eq:signn}
  &-C_1\Re \left ( H -i\epsilon i [H,A] -\zeta \right)
-
\epsilon^{-1} C_2 \Im \left ( H- i\epsilon i [H,A] -\zeta \right) \nonumber\\
&=
p^2 + C_2 W + V + g_\zeta(\epsilon);\;g_\zeta(\epsilon)=C_1\Re \zeta +\epsilon^{-1} C_2\Im \zeta. 
\end{align}
From the hypothesis on $W$ and $ V$, we have 
$ C_2 W + V \geq \langle x \rangle^{-\mu} $ for $C_2$ sufficiently
big. Fix such $C_2$. Next, notice that there exists a positive
$\epsilon_0^\prime$ such that for all 
$\zeta$ obeying \eqref{eq:conditions_on_epsilon_and_z} and for all
$\epsilon\leq\epsilon_0^\prime$ indeed $|\zeta|
\leq |\Re \zeta |+|\Im \zeta |\leq g_\zeta(\epsilon)$. Therefore, we can estimate
\begin{align*}
  T^* T \leq -C_1 \Re ( B^* R_{\zeta}(\epsilon)^* B) - 
C_2\epsilon^{-1} \Im ( B^* R_{\zeta}(\epsilon)^* B),
\end{align*}
yielding \eqref{eq:quadratic}.
\end{proof}

Notice that Lemma \ref{lem:quadratic} also holds with $\gamma$
replaced by $f_{|\zeta|}, \langle x \rangle^{-\mu/2}, p_j$ or
$|p|$. This follows from the proof since, for instance $\langle x
\rangle^{-\mu} \leq \gamma^2$.

\begin{lemma}
\label{lem:derivative}
The operator $R_{\zeta}(\epsilon)$ (defined in \eqref{eq:def_R_zeta})
has the following derivative,
$$
\frac{d}{d\epsilon} R_{\zeta}(\epsilon) =
(1-2i\epsilon)^{-1}
\left\{
R_{\zeta}(\epsilon)A -AR_{\zeta}(\epsilon)  +i\epsilon
R_{\zeta}(\epsilon) (x \cdot \nabla W) R_{\zeta}(\epsilon)
\right\}.
$$
\end{lemma}

\begin{proof}
We compute
\begin{align*}
& \frac{d}{d\epsilon} R_{\zeta}(\epsilon)  =
- R_{\zeta}(\epsilon) [H,A] R_{\zeta}(\epsilon) \\
&=  R_{\zeta}(\epsilon) A - A R_{\zeta}(\epsilon) + \epsilon
R_{\zeta}(\epsilon) [[ H, A],A] R_{\zeta}(\epsilon) \\
&=  R_{\zeta}(\epsilon) A - A R_{\zeta}(\epsilon)
-2i\epsilon R_{\zeta}(\epsilon) [ H, A] R_{\zeta}(\epsilon) +
i\epsilon R_{\zeta}(\epsilon) (x\cdot \nabla W) R_{\zeta}(\epsilon).
\end{align*}
We now insert the first line of the above calculation in the
last and isolate $\frac{d}{d\epsilon} R_{\zeta}(\epsilon)$ in the
resulting equation.
\end{proof}

Using Mourre's technique of differential inequalities, we
can (with some work) prove limiting absorption principle estimates.

\begin{lemma}
\label{lem:TheHardPart}
For all $\epsilon_0>0$ there exists $\epsilon_0^\prime>0$ such that
for all positive 
$\epsilon\leq \epsilon_0^\prime$, all positive $\delta<1/2$ and all $\zeta$ as in
\eqref{eq:conditions_on_epsilon_and_z}  we have 
the following estimate uniformly
in $\zeta$ and $\epsilon$, with $f=f_{|\zeta|}$ given by \eqref{eq:def_f},
\begin{align}
\label{eq:weighted_f_5.1}
\Big\| \langle \epsilon A \rangle^{-1} f_{|\zeta|}^{1/2}
\langle x \rangle^{-(1/2+\delta)} R_{\zeta}(\epsilon)
\langle x \rangle^{-(1/2+\delta)} f_{|\zeta|}^{1/2}
\langle \epsilon A \rangle^{-1} \Big\| \leq
C_{\delta}.
\end{align}
\end{lemma}

\begin{proof}
We study the function
$$
F_{\zeta}(\epsilon) = \langle \epsilon A \rangle^{-1} f^{1/2}
\langle x \rangle^{-s} R_{\zeta}(\epsilon) \langle x
\rangle^{-s} f^{1/2} \langle
\epsilon A
\rangle^{-1};\;s=1/2 + \delta.
$$
From Lemma \ref{lem:quadratic} we get
\begin{equation}
\label{eq:a_priori}
\| F_{\zeta}(\epsilon) \| \leq C \epsilon^{-1}.
\end{equation}
We will prove
\begin{equation}
\label{eq:diff_ineq}
\| \frac{d}{d\epsilon} F_{\zeta}(\epsilon) \| \leq
C \left\{ \|  F_{\zeta}(\epsilon) \| +
\epsilon^{-1+\delta'} \|  F_{\zeta}(\epsilon) \|^{1/2}
\right\},
\end{equation}
for any positive $\delta' \leq \delta$.
Combining \eqref{eq:a_priori} and \eqref{eq:diff_ineq}
we clearly get \eqref{eq:weighted_f_5.1} by repeated integration.

To prove \eqref{eq:diff_ineq} we calculate
$\frac{d}{d\epsilon} F_{\zeta}(\epsilon)$ using Lemma
\ref{lem:derivative}. After commutation we need to estimate
the following terms (and some `adjoint' expressions):

\begin{align}
\label{eq:first_term}
&
\left\|
 \langle \epsilon A \rangle^{-1} Af^{1/2}
\langle x \rangle^{-s} 
R_{\zeta}(\epsilon) \langle x
\rangle^{-s} f^{1/2} \langle
\epsilon A
\rangle^{-1}\right\|, \\
\label{eq:second_term}
&
\left\|
\langle \epsilon A \rangle^{-3} \epsilon A^2f^{1/2}
\langle x \rangle^{-s} 
R_{\zeta}(\epsilon) \langle x
\rangle^{-s} f^{1/2} \langle
\epsilon A
\rangle^{-1}\right\|, \\
\label{eq:third_term}
&\;
\epsilon \Big\|\langle \epsilon A \rangle^{-1} f^{1/2}
\langle x \rangle^{-s} 
R_{\zeta}(\epsilon) \nonumber \\
&
\,\,\,\,\,\,\,\,\,\,\,\,\,\,\,\,\,\,\,\,
\,\,\,\,\,\,\,\,\,\,\,\,\,\,\,\,\,\,\,\,
\times
(x\cdot \nabla W) 
R_{\zeta}(\epsilon) \langle x
\rangle^{-s} f^{1/2} \langle
\epsilon A
\rangle^{-1}\Big\|.
\end{align}
Here and below it is useful to note the uniform bound
\eqref{eq:bounda} which is also valid with the present definition of $f_E$. 

The term $x\cdot \nabla W$ of \eqref{eq:third_term} satisfies
$|x\cdot \nabla W| \leq C \langle x \rangle^{-\mu}$. Whence, using Lemma
\ref{lem:quadratic}, the
expression
\eqref{eq:third_term}
may be estimated in agreement with \eqref{eq:diff_ineq}.

So it suffices to
 prove that 
\begin{align}
\label{eq:sufficient_for_diff_ineq}
  \|A F_{\zeta}(\epsilon) \| \leq
C \epsilon^{-1+\delta'} \|  F_{\zeta}(\epsilon) \|^{1/2}.
\end{align}
Using that 
$$
\|\langle \epsilon A \rangle^{-1} \langle A
\rangle^{1/2-\delta'} \| \leq C \epsilon^{-1/2+ \delta'},
$$
\eqref{eq:sufficient_for_diff_ineq} will follow from 
the bound
\begin{equation}
\label{eq:neces_for_diff_eq_new}
\| \langle A \rangle^{1/2 + \delta'} f^{1/2} \langle x \rangle^{-s}
R_{\zeta}(\epsilon) \psi \|^2 \leq C \epsilon^{-1}
\|F_{\zeta}(\epsilon)\|,
\end{equation}
where $\psi = f^{1/2} \langle x \rangle^{-s} \langle \epsilon A
\rangle^{-1} \phi$; $\| \phi\| = 1$. For this bound  we estimate 
\begin{align*}
&\left( \langle A \rangle^{1/2 + \delta'} f^{1/2}\langle x
\rangle^{-s} \right)^* \langle A \rangle^{1/2 + \delta'}
\langle x \rangle^{-s} f^{1/2}\\
&\leq 
C_1 f \langle x \rangle^{-2s} +
f^{1/2} \langle x \rangle^{-s} A^2 \langle A \rangle^{2 \delta' -1}
\langle x \rangle^{-s} f^{1/2} \\
&\leq 
\sum_{j=1}^d \Re
\left (
p_j  B_j\langle A \rangle^{2 \delta'}f \langle x \rangle^{-2\delta} \right )+ C_2f^{1/2}\langle x \rangle^{-s}\langle A \rangle^{2 \delta'}
\langle x \rangle^{-s} f^{1/2};\\
&B_j=f^{1/2}\langle x \rangle^{-s}x_j\langle A \rangle^{2 \delta'-1}Af^{-1/2}\langle x \rangle^{s-1}\langle A \rangle^{-2 \delta'}.
\end{align*}

 The operators $B_j$ are bounded uniformly in
 $\zeta$. This may readily be proved by the techniques of the proof of
 \cite[Lemma
 2.5]{MoSk} and will not be done  here. (We move the factor 
 $\langle A \rangle^{2 \delta'-1}$ to the right by commutation using
 the representation formula \cite[(2.5)]{MoSk}.)

The two  terms on the right hand side are estimated in the same
fashion in  the expectation in
the state $R_{\zeta}(\epsilon) \psi$, so let us only consider the
first one in details.
 We get with $g=\langle x
\rangle^{-2\delta} f$ 
\begin{align*}
& \left|
\langle p_j R_{\zeta}(\epsilon) \psi,  B_j\langle A \rangle^{2 \delta'}gR_{\zeta}(\epsilon) \psi \rangle
\right| \\
&\leq C
\left\| p_j R_{\zeta}(\epsilon) f^{1/2}\langle x \rangle^{-s} \langle
\epsilon A \rangle^{-1} \right\|
\cdot
\left\|  \langle A \rangle^{2 \delta'} gR_{\zeta}(\epsilon) \langle x
\rangle^{-s} f^{1/2}\langle \epsilon A
\rangle^{-1} \right\|.
\end{align*}
An application of Lemma \ref{lem:quadratic} to the first
factor gives
$$
\left\| p_j R_{\zeta}(\epsilon) f^{1/2}\langle x \rangle^{-s} \langle
\epsilon A \rangle^{-1} \right\|
\leq
C \epsilon^{-1/2} \| F_{\zeta}(\epsilon) \|^{1/2}.
$$

To estimate the second factor we use \cite[Lemma 7.1]{MoSk}, which  says that
for all $s^\prime>0$ there exists a constant $C>0$ such that
$$
\| \langle A \rangle^{s^\prime} \langle x \rangle^{-s^\prime} \langle p \rangle^{-s^\prime} \|
\leq C,
$$ to
write 
\begin{align*}
&\left\| \langle A \rangle^{2 \delta'} g_jR_{\zeta}(\epsilon) \langle x
\rangle^{-s} f^{1/2}\langle \epsilon A
\rangle^{-1} \right\| \\
&\leq
C \left\| \langle p \rangle\langle x
\rangle^{2\delta'} g R_{\zeta}(\epsilon) \langle x
\rangle^{-s} f^{1/2}\langle \epsilon A
\rangle^{-1} \right\|.
\end{align*}
By Lemma
\ref{lem:quadratic}
\begin{align*}
\left\| \langle x
\rangle^{2\delta'} g
R_{\zeta}(\epsilon)f^{1/2}
\langle x \rangle^{-s} \langle \epsilon A \rangle^{-1}
\right\| \leq C \epsilon^{-1/2} \|F_{\zeta}(\epsilon) \|^{1/2}.
\end{align*}
Moreover 
\begin{align*}
&(p_i \langle x
\rangle^{2\delta'} g)^*p_i  \langle x
\rangle^{2\delta'} g
=
p_i g^2\langle x\rangle^{4\delta'}
p_i  +O(\langle
x\rangle^{-2+4\delta'-4\delta}f^2)\\
&=
p_i O(1)p_i  +O(f^2),
\end{align*}
from which we obtain  the  final estimate 
\begin{align*}
\left\| \langle A \rangle^{2 \delta'} gR_{\zeta}(\epsilon) \langle x
\rangle^{-s} f^{1/2}\langle \epsilon A
\rangle^{-1} \right\| \leq C \epsilon^{-1/2} \|F_{\zeta}(\epsilon)
\|^{1/2}
\end{align*}
by another application of Lemma \ref{lem:quadratic}.
Therefore we have proved that  the contribution from the first term
agrees with 
the bound 
\eqref{eq:neces_for_diff_eq_new}. Since the  other
term can be treated similarly the proof of \eqref{eq:weighted_f_5.1} 
is done.
\end{proof}

By taking $\epsilon\to 0$ we obtain

\begin{cor}
\label{cor:unperturbed}
For any $\delta > 0$ and $\theta \in (0,\pi)$ the operators
$$
\zeta \mapsto \langle x \rangle^{-(1/2+\mu/4+\delta)}
(H-\zeta)^{-1} \langle x \rangle^{-(1/2+\mu/4+\delta)},
$$
 and 
$$
\zeta \mapsto \langle x \rangle^{-(1/2+\delta)}
f_{|\zeta|}^{1/2} (H-\zeta)^{-1} f_{|\zeta|}^{1/2} 
\langle x \rangle^{-(1/2+\delta)}
$$ are bounded uniformly 
in the sector 
$\Gamma_{\theta}$.

\end{cor}
 
\begin{remark}
\label{rem:hoelder}
One may show uniform H{\"o}lder continuity in
$\Gamma_{\theta}$ of the first function in Corollary
\ref{cor:unperturbed}
by a certain modification of the Mourre type method of \cite{PSS}
using 
 bounds from the present section.
The proof we give in Section \ref{iterated_perturbed} is more
efficient
and yields a better H{\"o}lder exponent although it requires
more smoothness of the potential.
\end{remark}

\section{Iterated resolvents for $V_1$}
\label{iterated}
\subsection{Discussion and statement of results}

In this section we will use a strategy similar to the one
from \cite{GIS} to prove weighted estimates for iterated
resolvents. In \cite{GIS} this strategy was developed to prove weighted estimates for
powers of resolvents at energies for which a Mourre estimate
holds. Although the different (multiple commutator) approach of \cite{JMP} can be
extended to give some results in the present context, it
seems impossible to derive {\it
optimal} weights by some \cite{JMP}-type technique. Only the case
$V=V_1$ is considered, as in  Section \ref{sec:lim_abs_unperturbed}.

The following symbols will play a  prominent role:
\begin{align}
\label{eq:def_a0_b}
a_0(x,\xi) = \frac{\xi^2}{f_E(x)^2} ,\;\;
b(x,\xi)  = \frac{x}{\langle x \rangle} \cdot \frac{\xi}{f_E(x)}.
\end{align}
Here and in the rest of the paper
\begin{equation}
  \label{eq:def_f_2}
  f=f_E=\sqrt{\kappa_0^{-2}E + (1-\mu/2)^{-1} \langle x
  \rangle^{-\mu}}
\end{equation}
is essentially the function from \eqref{eq:def_f}---we have only introduced a
dependence on the constants $\mu, \kappa_0 > 0$ that appear  in Theorem
\ref{thm:resolvent_basic} below.
These constants are imposed on us, due to the need for
\eqref{postq}  and \eqref{eq:classgder} (and their quantum
analogues) to hold. In a part of this section the simpler definition
\eqref{eq:def_f} of $f$ would be sufficient, but in the critical
Subsection 
\ref{pseudodiff_Fefferman_Phong} the definition \eqref{eq:def_f_2}
above will be important.
These symbols will be studied together with the resolvent $R(\zeta)$. We
always take $E=|\zeta|$ with $\zeta \in \Gamma_{\theta}$ in \eqref{eq:def_a0_b}. Thus $a_0$ and $b$
depend on $\zeta$, even though we do not include this
explicitly in the notation. We also define a useful auxiliary weight
function $w$ and the symbol $h$ of the operator $H$ by
\begin{align}
  \label{eq:definitions}
  w = w_{E}(x) = \langle x \rangle f_{E}(x),  \;\;
  h=h(x,\xi) = \xi^2 + V(x).
\end{align}

We will let $\Op(a)$ denote the Weyl quantization of a 
symbol $a$. Explicitly $\Op(a)$ acts as follows
$$
(\Op(a) \phi)(x) = (2\pi)^{-d} \iint e^{i(x-y)\xi}
a((x+y)/2,\xi) \phi(y) \,dyd\xi,
$$
with mapping properties depending, of course, on the
symbol $a$.

We will be concerned with proving the following result:

\begin{thm}
\label{thm:resolvent_basic}
Let $V(x)$
satisfy the conditions of Theorem \ref{thm:lap} with
$V_2=0$. We reformulate the assumption (\ref{it:assumption3})
as: For some $\kappa_0 > 0$ and $2>\mu>0$,
\begin{align}
\label{eq:estim_W}
W(x) & =-2V(x)-x \cdot \nabla V(x) \geq 2\kappa_0^2 \langle x
\rangle^{-\mu}. 
\end{align} 
Let $\theta \in (0,\pi)$ and 
$\Gamma_{\theta}$ be as defined in \eqref{eq:def_Gamma_theta}. Let
$a_0$ and $b$ be as defined in \eqref{eq:def_a0_b}  with $E=|\zeta|$. Define $k=k(x)=\langle x \rangle^{1+\mu/2}$. 
\begin{subequations}
\label{group1}
Then the following conclusions, {\rm (\ref{it:one}) -
(\ref{it:last})}, hold for
$H = p^2 + V$ with all bounds being uniform in $\zeta \in \Gamma_{\theta}$:
\begin{enumerate}[\normalfont (i)]
\item \label{it:one} For all $\epsilon > 0$ 
there exists
$C>0$ such that
\begin{align}
  \label{eq:x_weights}
\|k^{-1/2-\epsilon} &R(\zeta) k^{-1/2-\epsilon} \| \leq C.
\end{align}
\item \label{item:C-0} There exists $C_0 > 0$, depending only on $V$,
 such that if
$F_{+} \in {\mathcal
    B}^{\infty}({\mathbb R})$, 
    $\supp (F_{+})\subset (C_0, \infty)$ and $F_{+}' \in
C_0^{\infty}({\mathbb R})$, then for all
$\epsilon > 0$
    and all $t > 0$ there exists $C>0$ such that 
\begin{align}\label{eq:quantum_part}
&\|k^{t -1/2-\epsilon} \Op(F_{+}(a_0)) R(\zeta) k^{-t-1/2-\epsilon} \| \leq C,
  \\
\label{eq:quantum_part222}
&\|k^{-t-1/2-\epsilon}R(\zeta)\Op(F_{+}(a_0)) k^{t -1/2-\epsilon} \| \leq C.
\end{align}
\item \label{some_label} Let
$
\tilde{F}_{+}, \tilde{F}_{-} \in {\mathcal B}^{\infty}({\mathbb R})$
satisfy (with $\kappa_0$ from \eqref{eq:estim_W}) for some
$\kappa > 0$,
\begin{itemize}
\item $
\inf \supp (\tilde{F}_{+}) >  - \kappa >
-\kappa_0,$ $
\sup
\supp
(\tilde{F}_{-}) <  \kappa < \kappa_0$. 
\item $\tilde{F}'_{-}, \tilde{F}'_{+} \in
C_0^{\infty}({\mathbb R})$.
\end{itemize}
 Let  $F_{-} \in C^{\infty}_{0}({\mathbb R})$. Then for all $\epsilon,
t> 0$ there exists $C > 0$ such that 
\begin{align}
  \label{eq:PsDO_part}
&\|k^{t -1/2-\epsilon} \Op(F_{-}(a_0)\tilde{F}_{-}(b))
R(\zeta) k^{-t-1/2-\epsilon}\| \leq C, \\
\label{eq:PsDO_part2}
&\|k^{-t-1/2-\epsilon} R(\zeta)
\Op(F_{-}(a_0)\tilde{F}_{+}(b)) k^{t
  -1/2-\epsilon}\| \leq C.
\end{align}
\item \label{it:last} Suppose $\tilde{F}_{+}$ and 
$\tilde{F}_{-}$ satisfy the
  assumptions from {\rm (\ref{some_label})}, $F_{-}^1,  F_{-}^2 \in C^{\infty}_{0}({\mathbb R})$
 and 
$$
\dist(\supp (\tilde{F}_{+}),\supp (\tilde{F}_{-})) > 0.
$$
Then for all $t> 0$ there exists $C > 0$ such that
\begin{align}
\label{eq:disjoint_b}
\|k^t \Op(F_{-}^1(a_0) \tilde{F}_{-}(b)) R(\zeta)
\Op(F_{-}^2(a_0)\tilde{F}_{+}(b)) k^t\| \leq C.
\end{align}
Suppose $F_{+}$ is given as in {\rm (\ref{item:C-0})},  some functions $\tilde{F}_{+},
\tilde{F}_{-}, F_{-}$  are given as in  {\rm (\ref{some_label})} 
 and suppose
$$\dist(\supp (F_{-}), \supp (F_{+})) > 0.$$
Then for all $t>0$
 there exists $C>0$
such that 
\begin{align}
  \label{eq:disjoint_a}
&\|k^t \Op(F_{+}(a_0)) R(\zeta) \Op(F_{-}(a_0)\tilde{F}_{+}(b)) k^t\| \leq C,\\
\label{eq:disjoint_a222}
&\|k^t \Op(F_{-}(a_0)\tilde{F}_{-}(b)) R(\zeta) \Op(F_{+}(a_0)) k^t\| \leq C.
\end{align}
\end{enumerate}
\end{subequations}
\end{thm}

Using purely algebraic arguments, cf. \cite{Jensen} or \cite{Isozaki},
one can get the following
result from Theorem
\ref{thm:resolvent_basic}. An elaboration is given in
Appendix \ref{algebra}. (See \eqref{eq:decom} for another
application of  the partition of unity needed in our case.) 

\begin{thm}
\label{thm:iterated_resolvents}
Let the assumptions and notations be as in Theorem
\ref{thm:resolvent_basic}.
\begin{subequations}
\label{group2}
Then the following conclusions hold for $H = p^2 + V$ (uniformly in $\zeta
\in \Gamma_{\theta}$):
\begin{enumerate}[\normalfont (i)]
\item Let $m \in {\mathbb N}$ and let $\epsilon > 0$ be
  arbitrary. Then there exists $C>0$ such that
\begin{align}
  \label{eq:x_weights_iterated}
\|k^{-(m-1/2)-\epsilon} R(\zeta)^m k^{-(m-1/2)-\epsilon} \| \leq C.
\end{align}
\item Let $C_0$ be the number from {\rm(\ref{item:C-0})} in Theorem
  \ref{thm:resolvent_basic} and suppose
  $F_{+}  \in {\mathcal B}^{\infty}({\mathbb R})$,
    $\supp (F_{+}) \subset (C_0, \infty)$ and
$F_{+}' \in C_0^{\infty}({\mathbb R})$.
    Then for all $m \in {\mathbb N}$
    and all $\epsilon, t > 0$ there exists $C>0$ such that
\begin{align}
  \label{eq:quantum_part_iterated}
&\| k^{t -1/2-\epsilon} \Op(F_{+}(a_0)) R(\zeta)^m
k^{-t-m+1/2-\epsilon}\| \leq C, \\
\label{eq:quantum_part222_iterated}
&\|k^{-t-m+1/2-\epsilon}R(\zeta)^{m}\Op(F_{+}(a_0))
k^{t-1/2-\epsilon} \| \leq C.
\end{align}
\item \label{some_label_iterated} Let
$ \tilde{F}_{+}, \tilde{F}_{-}, F_{-}$ satisfy the assumptions from
{\rm (\ref{some_label})} in Theorem \ref{thm:resolvent_basic}.
Then for all $m \in {\mathbb N}$ and all  $\epsilon,
t> 0$ there exists $C > 0$ such that
\begin{align}
  \label{eq:PsDO_part_iterated}
&\|k^{t -1/2-\epsilon} \Op(F_{-}(a_0)\tilde{F}_{-}(b))
R(\zeta)^m k^{-t-m+1/2-\epsilon}\| \leq C, \\
\label{eq:PsDO_part2_iterated}
&\|k^{-t-m+1/2-\epsilon} R(\zeta)^m
\Op(F_{-}(a_0)\tilde{F}_{+}(b)) k^{t
  -1/2-\epsilon}\| \leq C.
\end{align}
\item Suppose $\tilde{F}_{+}$ and 
$\tilde{F}_{-}$ satisfy the
  assumptions from {\rm (\ref{some_label})}, $F_{-}^1,  F_{-}^2 \in C^{\infty}_{0}({\mathbb R})$
 and 
$$
\dist(\supp (\tilde{F}_{+}),\supp (\tilde{F}_{-})) > 0.
$$
Then for all $m \in {\mathbb N}$ and all  $t> 0$ there exists $C > 0$
such that
\begin{align}
\label{eq:disjoint_b_iterated}
\|k^t \Op(F_{-}^1(a_0) \tilde{F}_{-}(b)) R(\zeta)^{m}
\Op(F_{-}^2(a_0)\tilde{F}_{+}(b)) k^t\| \leq C.
\end{align}
Suppose $F_{+}$ is given as in {\rm (\ref{item:C-0})},  some functions
$\tilde{F}_{+},
\tilde{F}_{-}, F_{-}$  are given as in  {\rm (\ref{some_label})}
 and suppose
$$\dist(\supp (F_{-}), \supp (F_{+})) > 0.$$
Then for all $m \in {\mathbb N}$ and all  $t>0$ there exists $C>0$
such that
\begin{align}
  \label{eq:disjoint_a_iterated}
&\|k^t \Op(F_{+}(a_0)) R(\zeta)^{m}
\Op(F_{-}(a_0)\tilde{F}_{+}(b)) k^t\| \leq C,\\
\label{eq:disjoint_a222_iterated}
&\|k^t \Op(F_{-}(a_0)\tilde{F}_{-}(b)) R(\zeta)^{m}
\Op(F_{+}(a_0)) k^t\| \leq C.
\end{align}
\end{enumerate}
\end{subequations}
\end{thm}

The remaining part of the present section is devoted to proving Theorem
\ref{thm:resolvent_basic}. 

The
first statement \eqref{eq:x_weights} is a restatement of
\eqref{eq:limitbound} (in the case $V_2=0$) which was proved in Section
\ref{sec:lim_abs_unperturbed}, cf. Corollary \ref{cor:unperturbed}.

In Subsection \ref{pseudodiff} we shall discuss a 
pseudodifferential calculus given in terms of the metric $\frac{dx^2}{\langle x \rangle^2}
+ \frac{d\xi^2}{f_E^2}$, where $f_E$ is the function from
\eqref{eq:def_f_2}.

The localization
$F_{+}$ in the second equation \eqref{eq:quantum_part} can
be thought of as an energy localization (uniform in energy). In
Subsection \ref{QM} we use the pseudodifferential calculus to deal
with \eqref{eq:quantum_part}, and with \eqref{eq:quantum_part222},
\eqref{eq:disjoint_a} and \eqref{eq:disjoint_a222}. 

The symbol
$F_{-}(a_0)\tilde F_{-}(b)$ lies in a good symbol class;
 this would not have been the case
without the factor
$F_{-}(a_0)$. Thus to prove \eqref{eq:PsDO_part} we can use
the  pseudodifferential calculus. A positive commutator  will play a major role in
the analysis, which is carried out in
Subsection \ref{pseudodiff_Fefferman_Phong}. The remaining estimates, \eqref{eq:PsDO_part2}
and \eqref{eq:disjoint_b}, can be proved similarly.

\subsection{Pseudodifferential calculus}
\label{pseudodiff}

We will use pseudodifferential operators in the Weyl calculus
associated with the metric $g_E = \frac{dx^2}{\langle x \rangle^2}
+ \frac{d\xi^2}{f^2}$ (cf. \cite[Chapt. XVIII]{H3}), where
$f=f_{E}$ is the energy-dependent function given in
\eqref{eq:def_f_2}.  For a part of the argument one could
instead use the (energy-independent) metric
$\frac{dx^2}{\langle x \rangle^2} +
\frac{d\xi^2}{\langle x \rangle^{-\mu}}$. However the crucial
positivity arguments in Subsections  \ref{QM} and \ref{pseudodiff_Fefferman_Phong}
(applications of the Fefferman-Phong inequality) rely indeed on 
the more precise energy-dependent estimates. It is clear that
for $E=0$, $f_E = C \langle x \rangle^{-\mu/2}$ and the two
metrics are essentially equal.

Since $\mu < 2$ we have a `Planck's constant' of size
$$
w^{-1} = \langle x \rangle^{-1} f^{-1} \leq 
(1-\frac{\mu}{2})^{1/2}  \langle x \rangle^{-1+\mu/2}.
$$ 

We will prove that $\frac{dx^2}{\langle x \rangle^2} +
\frac{d\xi^2}{f^2}$ satisfies the definition of a H\"{o}rmander metric
with estimates that are uniform in $E \in \left( 0,1\right]$. Therefore we get
uniform (in $E$) control of the constants appearing in the
pseudodifferential calculus. In particular the boundedness
(on $L^2({\mathbb R}^d)$) of pseudodifferential
operators (\cite[Theorem 18.6.3]{H3}) and the positivity (to highest order) of pseudodifferential
operators with positive symbols (the Fefferman-Phong inequality
\cite[Theorem 18.6.8]{H3}) hold uniformly in $E$.

\begin{lemma}
\label{lem:hormander_metric} $\,$\\
\begin{enumerate}[\normalfont (i)]
\item (A uniform H\"{o}rmander metric $g_E$.)\\
For points $v \in {\mathbb R}^d \times {\mathbb R}^d$ write $v = (v_x,v_{\xi})$.
Define for $E\in \left( 0,1\right]$ and $(x,\xi) \in {\mathbb R}^d \times {\mathbb
  R}^d$ the metric $g(v)=g_E(v)=g_{(x,\xi)}(v)$ by
$$
g_{(x,\xi)}(v) = \frac{v_x^2}{\langle x \rangle^2} +
\frac{v_{\xi}^2}{f_E^2(x)}.
$$
Then $g$ is a H\"{o}rmander metric uniformly in $E \in \left( 0,1\right]$, i.e.
there exist constants $C_1, C_2, N > 0$ independent of $E \in \left( 0,1\right]$ such that $g$
satisfies
\begin{itemize}
\item {\rm (slow variation)} If $g_{(x,\xi)}((y,\eta)) \leq
  1/C_1$ then
$$
 g_{(x,\xi)+(y,\eta)}(v) \leq C_1 g_{(x,\xi)}(v) \text{ for
  all }v \in {\mathbb R}^d \times {\mathbb R}^d.
$$
\item {\rm (uncertainty principle)} $g_E \leq g_E^{\sigma}$, where
  $g_E^{\sigma}$ denotes the dual metric of $g_E$ with respect to the
  standard symplectic form $\sigma = dx \wedge d\xi$.
\item {\rm (temperateness)} For all $v \in {\mathbb R}^d \times
{\mathbb
    R}^d\setminus \{0\}$  and all
$(x, \xi), (y, \eta) \in {\mathbb R}^d \times {\mathbb R}^d$ we have
$$
g_{(x,\xi)}(v)\leq C_2g_{(y,\eta)}(v)
\left( 1 + g_{(x,\xi)}^{\sigma}((y,\eta)-(x,\xi))\right)^N.
$$
\end{itemize}
\item (Uniform weight functions.)\\
A positive function $m=m_E=m(x,\xi)$ is said to be a uniformly temperate weight
(w.r.t.  $\sigma, g$) if the following two
conditions are satisfied with constants
independent of $E\in \left( 0,1\right]$
\begin{itemize}
\item There exists $c, C > 0$ such that for all $v, v_1 \in
{\mathbb R}^{2d}$: 
$$
g_v(v_1)
\leq c
\Rightarrow m(v)/C \leq  m(v+v_1) \leq C m(v).
$$
\item There exists $C, N > 0$ such that for all $v,v_1 \in
{\mathbb R}^{2d}$: 
$$
m(v_1) \leq C m(v) \left( 1 +
g_{v_1}^{\sigma}(v-v_1)\right)^N.
$$
\end{itemize}
With this definition, any of the functions $\langle x \rangle,\;
f_E,\;\langle \xi \rangle,\;$or $\langle \frac{\xi}{f_E} \rangle$, as
well as any combination of products of real powers of these functions,  is a uniformly temperate weight
function for the metric $g_E$.
\end{enumerate}
\end{lemma}

The proof of Lemma \ref{lem:hormander_metric} is given in
Appendix
\ref{verification}. 

For a uniformly temperate weight function $m=m_E$  
we denote by $S_{\it {unif}}\left(m, g_E\right)$ the space
of $C^{\infty}$ functions (`symbols') $a=a_\zeta$ satisfying
\begin{align}
\label{eq:symbolclass}
|\partial_x^{\alpha} \partial_{\xi}^{\beta} a(x,\xi)| \leq C_{\alpha,
  \beta}  m(x,\xi)\langle x \rangle^{-|\alpha|}f_E^{-|\beta|},
\end{align}
for all $\alpha, \beta \in ({\mathbb N}\cup\{0\})^d$ with constants
$C_{\alpha,\beta}$ independent of $\zeta =Ee^{i\phi}\in \Gamma_\theta$. We
let $\Psi_{\it {unif}}\left(m, g_E\right)$ denote the space of
operators given as the Weyl quantization of symbols from 
$S_{\it {unif}}\left(m,g_E\right)$. 
Thus, for instance, a short verification gives (with $h$ from
\eqref{eq:definitions}) 
\begin{align}
\label{eq:symbolclass_h}
h,h-\zeta \in
S_{unif}\left(f_{|\zeta|}^2\langle \frac{\xi}{f_{|\zeta|}}\rangle^2,
g_{|\zeta|}\right).
\end{align}

Notice  that for any $C_0^{\infty}$--function
$G$ we have with $b$ from \eqref{eq:def_a0_b} for $E=0$,
$$
\partial_x^{\alpha} \partial_{\xi}^{\beta} G(b) \approx (\langle x
\rangle^{(\mu/2-1)}|\xi|)^{|\alpha|}\langle x \rangle^{\mu |\beta|/2}+\cdots.
$$
Consequently $G(b)$ is not a good symbol (not even with $\xi$
considered as bounded since $\mu$ can be greater than  $1$). The
remedy for this has already been introduced in \eqref{group1}: Let
$F_{-} \in C_0^{\infty}({\mathbb R})$ and study $F_{-}(a_0) G(b)$ for
any $C^{\infty}$--function $G$. Notice that $|b|^2 \leq a_0$, so
$b$ is bounded on $\supp (F_{-}(a_0))$. Using the elementary
bound, valid for any $s\in{\mathbb R}$,
\begin{equation}
    \label{eq:bounda}
    |\partial_x^{\alpha}f_E^s|\leq C_{\alpha}f_E^s\langle x
\rangle^{-|\alpha|},
  \end{equation}
with $C_{\alpha}$ independent of $E$ 
(or in short $f_E\in S_{\it {unif}}\left( f_E, g_E\right)$ recalling
the convention $E=|\zeta|$), we readily infer that
indeed  
$$F_{-}(a_0) G(b) \in S_{\it {unif}}\left(1,g_E\right).$$

Once Lemma \ref{lem:hormander_metric} is established we 
have from \cite[Sections 18.4-6]{H3} (cf. in particular \cite[Theorems 18.6.3 and
18.6.8]{H3}): 
 
\begin{thm}
\label{thm:CV_and_FPh_uniform} $\,$\\
\begin{enumerate}[\normalfont (i)]
\item \label{L^2bound} Let $a \in S_{\it {unif}}\left(1,g_E\right)$. Then there exists a constant $C > 0$, independent of
$\zeta \in \Gamma_\theta$, such that
$$
\| \Op (a) \| \leq C.
$$
\item \label{Lowbound}Let $a \in S_{\it {unif}}\left(w_E^2
,g_E\right)$ and suppose $a\geq 0$. Then there exists
a constant $C > 0$ independent of
$\zeta \in \Gamma_\theta$ such that
$$
\Op (a) \geq -C,
$$
as a quadratic form on ${\mathcal S}({\mathbb R}^d)$.
\end{enumerate}
\end{thm}

Another useful tool is the composition
rule for pseudodifferential operators (see \cite[Theorem 18.5.4]{H3}): Suppose  $a_1\in 
S_{\it {unif}}\left( m_1,
g_E\right)$ and $a_2\in 
S_{\it {unif}}\left( m_2,
g_E\right)$. Then $\Op(a_1)
\Op(a_2) = \Op(s)$ with $s\in 
S_{\it {unif}}\left( m_1m_2,
g_E\right)$, and asymptotically $s = \sum_{j=0}^\infty s_j(x,\xi)$ 
with $$s_j(x,\xi)\in 
S_{\it {unif}}\left( w^{-j}_Em_1m_2,
g_E\right)$$  given by
\begin{align}
s_j(x,\xi) = 2^{-j} i^j\sum_{|\alpha + \beta| = j}
\frac{(-1)^{|\alpha|}}{\alpha ! \beta !}
(\partial_{\xi}^{\alpha} \partial_x^{\beta} a_1)
(\partial_{\xi}^{\beta} \partial_x^{\alpha} a_2). 
\end{align} 
Here `asymptotically' means that for all $N \in
{\mathbb N}$ we have
$$
s-\sum_{j=0}^N s_j\in S_{\it {unif}}\left(w^{-(N+1)}_E, g_E
\right).
$$

Finally we shall discuss some further results for the class $S_{\it
  {unif}}\left(m,g_E\right)$. Although they will not be used in this
  paper, we feel that including them might be clarifying for the reader.

Let us topologize  $S_{\it {unif}}\left(m,g_E\right)$
by the seminorms $\|a\|_{\alpha,\beta}$, each defined as the smallest
constant $C_{\alpha, \beta}$ independent of $\zeta=Ee^{i\phi}\in \Gamma_\theta$ such that
\eqref{eq:symbolclass} holds.  From \cite[Theorem 18.5.10]{H3} (and its proof) we
learn that 
\begin{align}
  \label{eq:gauss} 
&e^{i\kappa \langle p_x,p_{\xi}\rangle}:S_{\it
  {unif}}\left(m,g_E\right)\to S_{\it {unif}}\left(m,g_E\right)
  \;\nonumber\\
&{\rm {is\;a \;topological \;isomorphism}};\;\kappa\in {\mathbb R}.
\end{align}
Of particular interest are the values
$\kappa=-1,-\frac{1}{2},\frac{1}{2},1$, cf. \cite[(18.5.20)]{H3},
linking Weyl quantization with left and right Kohn-Niren\-berg
quantization.  

Yet another basic result is a uniform (partial) version of the Beals
criterion \cite[Theorem 4.4]{Beals} (see also \cite{BoCh}). It is  a useful tool for linking pseudodifferential
and functional calculi, cf. \cite[Appendix D]{DeGe}. 
We introduce  $\rm{ad}_B (A) = AB-BA$ and similarly for vector-valued operators,
$\rm{ad}_B^{\beta} (A)$, defined as $|\beta|$ compositions of
operations of the previous type. The Beals criterion in our case is the characterisation of 
the space $\Psi_{\it{unif}}\left(1, g_E\right)$ as the set of 
$\mathcal B (L^2)$-valued functions $A=A_\zeta$ on $\Gamma_\theta$ for which for all multiindices $\alpha,
\beta \in {\mathbb N}^d$ 
\begin{equation*}
\|\langle x\rangle^{|\alpha|}f_{|\zeta|}^{|\beta|}{\rm{ad}}^{\alpha}_p{\rm{ad}}^{\beta}_x(A)\|
  \leq D_{\alpha,\beta}, 
\end{equation*} where the constants $D_{\alpha,\beta}$ are independent
  of $\zeta\in \Gamma_\theta$. This characterisation
may be proved using 
\eqref{eq:bounda}, \eqref{eq:gauss}, the proof of \cite[Theorem 1.4]{Beals77} and conjugation by the Fourier
transform, see the proof of \cite[Theorem D.8.2]{DeGe}.

\subsection{Energy estimate}
\label{QM}

In this subsection we will prove \eqref{eq:quantum_part}.
The proof of
\eqref{eq:quantum_part222} is very similar (we may proceed in the same
fashion for the
`adjoint' expression), and  the statements \eqref{eq:disjoint_a} and
\eqref{eq:disjoint_a222} may be proved along the same pattern using in
addition \eqref{eq:PsDO_part} and \eqref{eq:PsDO_part2} (the latter
estimates will be proved independently in Subsection
\ref{pseudodiff_Fefferman_Phong}).

We will start by proving the following lemma in which $f=f_{|\zeta|}$
and $w=w_{|\zeta|}$ are given by \eqref{eq:def_f_2} and 
\eqref{eq:definitions}, respectively.

\begin{lemma}
\label{lem:auxiliary}
For all $\epsilon > 0$, $s \geq 0$  and  all functions $F_{+}$ as in
\eqref{eq:quantum_part} the operator
\begin{equation}
  \label{eq:F_+(a0)}
 \langle x \rangle^{-1/2-\epsilon} f^{1/2}w^s \Op( F_{+}(a_0)) R(\zeta) w^{-s}
f^{1/2} \langle x \rangle^{-1/2-\epsilon}, 
\end{equation}
is bounded uniformly in $\zeta \in \Gamma_{\theta}$.
\end{lemma}

\begin{proof}
 
We will use the inequality (with $h$ from \eqref{eq:definitions}),
\begin{align}
\label{eq:1eqa}
\frac{\xi^2}{f^2} =\frac{\Re(h-\zeta)}{f^2} -\frac{\Re(V-\zeta)}{f^2} \leq \frac{\Re(h-\zeta)}{f^2} + C,
\end{align}
where the constant $C$ only depends on $V$.

By \eqref{eq:1eqa}
$$
F_{+}(a_0)^2 \leq F_{+}(a_0)^2 \frac{\xi^2}{f^2 C_0} \leq F_{+}(a_0)^2
\frac{f^{-2} \Re(h-\zeta) + C}{C_0}.
$$
Thus, if $C/C_0\leq 1/2$, i.e. if $C_0$ has been chosen sufficiently
large, 
\begin{align}
\label{eq:symbol_ineq}
F_{+}(a_0)^2 
\leq  
\frac {2}{C_0}  F_{+}(a_0)^2 \frac{\Re(h-\zeta)}{f^2}.
\end{align}
By \eqref{eq:symbolclass_h} and \eqref{eq:bounda} this is an inequality for symbols in $S_{unif}(\langle \frac{\xi}
{f}\rangle^2, g_{|\zeta|})$. The constant $C_0 > 0$ only  depends on
$V$ as demanded in Theorem \ref{thm:resolvent_basic} (\ref{item:C-0}).

For $s=0$, the bounds of \eqref{eq:F_+(a0)}  follow from Corollary
\ref{cor:unperturbed} and the
pseudodifferential calculus.

Suppose we have proved uniform boundedness (in $\zeta$) of the
operator \eqref{eq:F_+(a0)}  
for all $\epsilon>0$, functions $F_{+}$ as in
\eqref{eq:quantum_part},  and $s \leq s_0$ for some $s_0 \geq 0$. We
will then prove that the operator
\begin{align*}
  \Op(f w^{s-1/2} F_{+}(a_0)) R(\zeta) f w^{-s-1/2-\epsilon}
\end{align*}
is also uniformly bounded for all $\epsilon>0$, such functions $F_{+}$,  and $s \in (s_0, s_0 +1
)$. Since
$$
f w^{s-1/2} = (\langle x \rangle^{-1/2}w^{-\delta})f^{1/2} w^{s+\delta},
$$
and $w = \langle x \rangle f \geq  \langle x \rangle^{1-\mu/2}$, this
is equivalent to uniform boundedness  of \eqref{eq:F_+(a0)}  
 for $s$ is
 the same range.   

Thus our  goal will be to prove a uniform bound on
$$
 \|\Op(f w^{s-1/2}F_{+}(a_0)) R T^{-1} \|,
$$
where $R=R(\zeta)$ and $T^{-1}=f w^{-s-1/2-\epsilon}$.

Using Theorem \ref{thm:CV_and_FPh_uniform},
\eqref{eq:symbol_ineq} and the symbolic calculus we may estimate 
\begin{align}
  \label{eq:fef10}
  &\Op(a)^*P\Op(a)\geq -C;\\
&a=\langle \frac {\xi} {f}\rangle^{-1}f^{-1}w^{\frac
  {3}{2}-s}\;(\in S_{unif}(a, g_{|\zeta|})),\nonumber\\
&P=\Op\left (w^{2s-1}F_+(a_0)^2(\frac
  {2}{C_0}\Re(h-\zeta)-f^{2})\right ).\nonumber  
\end{align}

To write \eqref{eq:fef10} in a more convenient form we first use the
 standard parametrix construction to find a symbol
 $a^{(m)}\in S_{unif}(a^{-1}, g_{|\zeta|})$ such that 
 \begin{equation*}
 \Op(a)\Op(a^{(m)})-I=\Op(r^{(m)})\in \Psi_{unif}(\langle x \rangle^{-m}, g_{|\zeta|});\; m>0. 
 \end{equation*}
Pick a function 
 ${F}_{+}^1$  satisfying  the same assumptions as $F_{+}$, and
furthermore, ${F}_{+}^1 = 1$ on a neighbourhood of $\supp
F_{+}$. We readily show, that for  any $m>0$
\begin{align}
  \label{eq:fef11}
  &P\geq -D^*D-C_1\Op(r_{m}^1);\\
&
D=\Op(\langle \frac {\xi} {f}\rangle)B_1,\;B_1=\Op(b_1{F}_{+}^1(a_0)),\;b_1\in S_{unif}(fw^{s-\frac {3}{2}},
g_{|\zeta|}),\nonumber\\
&r_{m}^1 \in S_{unif}(\langle \frac {\xi}
  {f}\rangle^2\langle x\rangle^{-m}, g_{|\zeta|}).\nonumber
\end{align}

Using \eqref{eq:fef11} we have to bound
\begin{multline}
\label{fef}
T^{-1} R^* 
\Big ( \frac
  {2}{C_0}\Op\left (w^{2s-1}F_+(a_0)^2\Re(h-\zeta)\right )\\
+D^*D+C_1\Op(r_{m}^1)\Big)R T^{-1}.
\end{multline}
For the contribution from the first  term in \eqref{fef}, we write

\begin{multline}
\label{calculus}
\Op\left (w^{2s-1}F_+(a_0)^2\Re(h-\zeta)\right )
\leq \Re \left( \Op\left (w^{2s-1}F_+(a_0)^2\right
  )(H-\zeta)
\right)\\
+\hat B_1^*\Op(\langle \frac {\xi} {f}\rangle)^2\hat B_1+C_2\Op(\hat r_{m}^1),
\end{multline}
where $\hat B_1$ and $\hat r_{m}^1$ have the same form as $B_1$ and
$r_{m}^1$, respectively.
Clearly we may estimate the contribution from the first term on the
right hand side of \eqref{calculus} by the  induction hypothesis since
\begin{align*}
&T^{-1} R^* \Re \left( \Op\left (w^{2s-1}F_+(a_0)^2\right
  )(H-\zeta)\right)
R T^{-1} \\
&=
\Re \left( T^{-1} R^* \Op(w^{2s-1}F_+(a_0)^2T^{-1} \right)\\
&= 
\Re \left( (\{\Op(a)f w^{s-3/2-\epsilon} \Op(F_{+}^1(a_0))+
  \Op(r_{m})\}R f
  w^{-s-1/2-\epsilon} )^*\right),
\end{align*}
where $a\in S_{unif}(1, g_{|\zeta|})$ and $r_{m}\in S_{unif}(\langle x \rangle^{-m}, g_{|\zeta|})$.

So in
order to finish the proof we only have to take care of the second and
third 
terms from \eqref{fef}. We write for the second one
\begin{equation}
  \label{eq:1junk}
D^*D=
 \Re \left( B_1^*\check B\Op(\langle \frac {\xi} {f}\rangle^2)\right)+\Op(\check r_{m}), 
\end{equation}
where $\check B =\Op(\check b \check{F}_{+}^1(a_0))$ with $\check b
  \in S_{unif}(fw^{s-\frac {3}{2}}, g_{|\zeta|})$ and 
  $\check {F}_{+}^1$ is given as ${F}_{+}^1$ but such that $\check {F}_{+}^1 = 1$ on a neighbourhood of $\supp
F_{+}^1$, and 
$\check r_{m}\in S_{unif}(\langle \frac {\xi} {f}\rangle^2\langle x \rangle^{-m},
  g_{|\zeta|})$.

We write \begin{align*}
 &\Op(\langle \frac {\xi} {f}\rangle^2)=f^{-2}p^2 +
  \Op( a);\\& a=1 -i\xi \cdot \nabla f^{-2}+4^{-1}\Delta f^{-2} \in S_{unif}(\langle \frac {\xi} {f}\rangle ,
  g_{|\zeta|}).\end{align*}

Substituted into \eqref{eq:1junk} this yields 
\begin{equation}
  \label{eq:11junk}
D^*D=
 \Re ( B_1^*\check Bf^{-2}p^2)
 +\Re (D^*\tilde B)+\Op(\tilde r_{m}), 
\end{equation}
where $\tilde B$ and $\tilde r_{m}$ are defined similarly.
Next we substitute $p^2=(H-\zeta)+(\zeta-V)$ and apply Cauchy-Schwarz
to the second term to the right in \eqref{eq:11junk}. After a
subtraction we conclude that
 \begin{equation}
  \label{eq:111junk}
D^*D\leq
 2\Re \left( B_1^*\check Bf^{-2}(H-\zeta)\right)
 +\bar B^*\bar B +\Op(\bar r_{m}),
\end{equation}
for yet another couple of similar operators $\bar B$ and $\Op(\bar r_{m})$.

We may treat the contribution from the first term on the right hand
side of \eqref{eq:111junk} as above. The second term is handled by the
induction  hypothesis. 
The third 
term is similar to the third term from \eqref{fef}. For these terms 
we 
use the  resolvent identity $R(z)=R(i)+(z-i)R(i)R(z)$ and the fact
that $\langle x
\rangle^{-m} p^2R(i)\langle x
\rangle^{m}$
is bounded; we need
$m$ sufficiently large.
\end{proof}

The estimate \eqref{eq:quantum_part} is an easy consequence of Lemma
\ref{lem:auxiliary} and Lemma \ref{lem:changing_weights} below. Lemma
\ref{lem:changing_weights} will also be useful in
Subsection \ref{pseudodiff_Fefferman_Phong}.

\begin{lemma}
\label{lem:changing_weights}
Suppose $A=A_\zeta$ is a $\mathcal B (L^2)$-valued function in
$\zeta \in \Gamma_\theta$ such that for all $\epsilon > 0$,
$s \geq 0$ the operator
\begin{align}
\label{eq:Erik_Lem4_reform}
\langle x \rangle^{-\epsilon} f w^{s-1/2} A
R(\zeta) w^{-s-1/2} f \langle x \rangle^{-\epsilon}
\end{align}
is uniformly bounded. Then also 
$$
k^{t-1/2 -\epsilon} A R(\zeta) k^{-t-1/2 -\epsilon},
$$
is  bounded uniformly  in  $\zeta \in \Gamma_{\theta}$ for all $\epsilon > 0$, $t \geq 0$.
\end{lemma}

\begin{proof}[Proof of Lemma \ref{lem:changing_weights}]
Let $1= F_1 + F_2$ be a sharp partition of unity on ${\mathbb R}$,
$\supp (F_1) = (-\infty,0]$, $\supp (F_2) = [0, +
\infty)$ and let us consider the two terms
\begin{align}
\label{eq:F1_part}
\langle x \rangle^{-\epsilon} k^{t-1/2} A R(\zeta) k^{-t-1/2}  
\langle x \rangle^{-\epsilon} F_1(|\zeta| - \langle x \rangle^{-\mu}) \\
\label{eq:F2_part}
\langle x \rangle^{-\epsilon} k^{t-1/2} A R(\zeta) k^{-t-1/2}  
\langle x \rangle^{-\epsilon} F_2(|\zeta| - \langle x \rangle^{-\mu}).
\end{align}

First we study \eqref{eq:F1_part}. On $\supp (F_1(|\zeta| - \langle x
\rangle^{-\mu}))$ we have $ |\zeta| \leq \langle x \rangle^{-\mu}$ and therefore
\begin{align}
\label{eq:F1_f_estimates}
\langle x \rangle^{-\mu/2}\leq f_{|\zeta|}(x) \leq C \langle x \rangle^{-\mu/2}
.
\end{align}
Let us choose $s$ such that
\begin{align}
\label{F1_choice_s}
(-t-1/2)(1+\mu/2) = -s-1/2-(-s+1/2)\mu/2,
\end{align}
 that is $s = t(1+\mu/2)/(1-\mu/2)$ (notice that $s\geq 0)$, and therefore
\begin{align}
\label{F1_consequence_s}
(t-1/2)(1+\mu/2) = s-1/2+(-s-1/2)\mu/2.
\end{align}
So (using \eqref{F1_consequence_s}) to the left of the resolvent in \eqref{eq:F1_part} we can
write
$$
k^{t-1/2} = w^{s-1/2}f \left( \frac{\langle x \rangle^{-\mu/2}}{f}\right)^{s+1/2}.
$$
We know, since $f \geq \langle x \rangle^{-\mu/2}$, that the
$(\cdot)^{s+1/2}$-term is bounded.

To the right in \eqref{eq:F1_part} we can
write using \eqref{F1_choice_s}
$$
k^{-t-1/2} = w^{-s-1/2} f\left( \frac{\langle x \rangle^{-\mu/2}}{f}\right)^{-s+1/2}.
$$
Now we  infer uniform boundedness of \eqref{eq:F1_part} from
\eqref{eq:Erik_Lem4_reform} and \eqref{eq:F1_f_estimates}.

The boundedness of \eqref{eq:F2_part} is more subtle. Here we cannot
convert all $f$'s to $\langle x \rangle^{-\mu/2}$'s. Instead we have
to compare some $f$'s to $|\zeta|^{1/2}$ and others to $\langle x
\rangle^{-\mu/2}$. Notice that on $\supp (F_2(|\zeta| - \langle x
\rangle^{-\mu}))$ we have
\begin{align}
\label{eq:ineq_F2}
\langle x \rangle^{-\mu/2} \leq f_{|\zeta|}(x) \leq C |\zeta|^{1/2}.
\end{align}
For $s=(1+\mu/2)t$ we have
\begin{align}
\label{eq:F2_choice_s}
 (t-1/2) (1+\mu/2)&= -\mu/4+s-1/2, \\
\label{eq:cgoice_s'}
(t+1/2)(1+\mu/2)&=  s+1/2+\mu/4 .
\end{align}
To the left in \eqref{eq:F2_part} we write, using \eqref{eq:F2_choice_s},
\begin{align*}
k^{t-1/2} = \left( \frac{|\zeta|^{1/2}}{f}\right)^{s} 
\left( \frac{\langle x \rangle^{-\mu/2}}{f}\right)^{1/2}
|\zeta|^{-s/2} f w^{s-1/2}.
\end{align*}
To the right in \eqref{eq:F2_part} we write
\begin{align*}
k^{-t-1/2} = \left( \frac{f}{|\zeta|^{1/2}}\right)^{s} 
\left( \frac{\langle x \rangle^{-\mu/2}}{f}\right)^{1/2}
|\zeta|^{s/2} f w^{-s-1/2}.
\end{align*}

Now we  infer uniform boundedness of \eqref{eq:F2_part} from
\eqref{eq:Erik_Lem4_reform} and \eqref{eq:ineq_F2}.
\end{proof}

\subsection{Classical mechanics}
\label{Classical mechanics}
The purpose of this subsection is twofold: We present various
notation and computations needed in the next subsection.
Secondly we show how one can construct a classical
`propagation observable' which yields a classical analogue
of Theorem \ref{thm:minimal velocity}. The more technical
material of Subsection \ref{pseudodiff_Fefferman_Phong} may
be viewed as being based on the classical proof
presented here.

Recall the definitions of $a_0$, $b$, $f$, $h$, $w$ and $W$ from
\eqref{eq:def_a0_b}, \eqref{eq:def_f_2}, \eqref{eq:definitions} and \eqref{eq:estim_W}.
For any $E\geq 0$ we define
\begin{align*}
v = v_E(x) = \kappa_0^{-2} E + \langle x \rangle^{-\mu},
\end{align*}
and compute
\begin{align}\label{gradientg}
\nabla f(x) &= \frac{-\mu}{2-\mu} f^{-1} \langle x \rangle^{-\mu -1} \frac{x}{ \langle x \rangle}, \nonumber\\
\nabla w(x)& = (\kappa_0^{-2} E + \langle x \rangle^{-\mu})\frac{x}{w} = v \frac{x}{w}.
\end{align}
Recall also the definition of the Poisson bracket:
\begin{align*}
\{a,b\}_P = \nabla_{\xi}a \cdot  \nabla_{x}b - \nabla_{x}a \cdot   \nabla_{\xi}b.
\end{align*}
With our definition \eqref{eq:definitions} of $h(x,\xi)$ we get for any symbol $a$,
$
\{h,a\}_P = 2 \xi\cdot \nabla_{x}a - \nabla V \cdot\nabla_{\xi}a.
$
Thus, with $b$ as in \eqref{eq:def_a0_b}, an elementary
calculation yields
\begin{align}
\label{eq:positive_classic}
\{h,b\}_P = w^{-1}\left( 2h+W(x) - 2 b^2v(x) \right).  
\end{align}

\begin{thm}\label{thm:classicalminimal velocity bound}
Let $\kappa_0>0$ be given as in \eqref{eq:estim_W}. Then for any
classical orbit $x(t)$ with energy $E\geq 0$
\begin{equation}
  \label{eq:clasminimal}
 \liminf_{|t|\to\infty} |tC|^{-(1+\frac{\mu}{2})^{-1}}|x(t)|\geq 1;\;C=\kappa_0\frac{2+\mu}{(1-\frac{\mu}{2})^{\frac{1}{2}}}.
\end{equation}
\end{thm}

\begin{proof} We shall only prove the bound for $t\to +\infty$. 
Let us
fix 
$0<\tilde\kappa<\kappa < \kappa' \leq \kappa_0$.
We pick a real-valued, decreasing, smooth function 
$\tilde{F}_{-}$ such that 
$$\tilde{F}_{-}(\tilde\kappa)=1,\;\tilde{F}_{-}(\kappa)=0\;{\rm
  {and}}\;\supp (\tilde{F}_{-}')
\subset (\tilde\kappa, \kappa),$$
 and consider the observable
\begin{align}
\label{eq:def_q2}
q=q(x,\xi) = w (\kappa'-b)\tilde{F}_{-}(b).
\end{align}
We claim that  $q$ has a non-positive derivative:
First we compute using \eqref{eq:positive_classic}:
\begin{align}\label{postq}
\frac{d}{dt}\tilde{F}_{-}(b)&=\left( 2E+W - 2 b^2v
 \right)w^{-1}\tilde{F}_{-}^{\prime}(b) \nonumber \\
 &\leq 2(\kappa_0^2-\kappa^2)vw^{-1}\tilde{F}_{-}^{\prime}(b)\leq 0.
\end{align}
The contribution from the derivative  of the other factors on the right
hand side of \eqref{eq:def_q2} is computed and estimated using \eqref{gradientg}
and \eqref{eq:positive_classic} as 
\begin{align}\label{eq:classgder}
\left(\frac{d}{dt}\left( w (\kappa'-b)\right)\right)\tilde{F}_{-}(b) 
&=\left(
  2vb(\kappa^{\prime}-b)-2E-W+2vb^2\right)\tilde{F}_{-}(b)\nonumber \\
&\leq -\left( 2E+W-2v\kappa\kappa^{\prime}\right)\tilde{F}_{-}(b)\nonumber\\
&\leq -2(\kappa_0^2-\kappa\kappa^{\prime})v\tilde{F}_{-}(b)\leq 0.
\end{align}

In particular, we infer that 
\begin{align}
  \label{eq:qdecreas}
  \frac{d}{dt}q\leq -2(\kappa_0^2-\kappa\kappa^{\prime})\langle x
 \rangle^{-\mu}\tilde{F}_{-}(b)\leq 0.
\end{align}

By integrating \eqref{eq:qdecreas} we obtain the uniform bound
$$q(T)+\int_0^T\langle x(t)
 \rangle^{-\mu}\tilde{F}_{-}(b(t))dt\leq C,$$
by which we will now prove that 
\begin{equation}
  \label{eq:subseq}
  \tilde{F}_{-}(b(t))\to 0\;{\rm{for}}\; t\to\infty.
\end{equation}
Notice that by \eqref{postq}, 
$\tilde{F}_{-}(b(t))\to c;$ so we need only to show that $c=0$. There
are two cases: 
 1) $x$ is bounded, or 2)
$|x(t)|\to \infty$ along some sequence $t=t_n\to\infty$. In Case 1)
we have $\tilde{F}_{-}(b(t_n))\to 0$ for some sequence
$t_n\to\infty$ (by the boundedness of the integral) yielding $c=0$. In Case 2) we learn from the
boundedness of $q$ that $\tilde{F}_{-}(b(t_n))\to 0$ yielding $c=0$ in
this case too. 

Finally, define
$$F(r)=\frac{1}{2}\int_1^r(\kappa_0^{-2}E+(1-\frac{\mu}{2})^{-1}s^{-\mu})^{-\frac{1}{2}}ds,$$
and compute 
$$\frac{d}{dt} F(\langle x
 \rangle)= b.$$
Combined with \eqref{eq:subseq} this yields
$$\frac{d}{dt} F(\langle x
 \rangle)\geq \tilde\kappa {\;\rm{for}}\;t\geq t_{\tilde\kappa}.$$
Whence, by integrating, 
$$\frac{1}{2}(1-\mu/2)^{\frac{1}{2}}(1+\mu/2)^{-1}\langle x(t)
 \rangle^{1+\frac{\mu}{2}}\geq \tilde\kappa t -C.$$
Since $\tilde\kappa$ can be taken arbitrarily close to $\kappa_0$, we
are done.
\end{proof}

\begin{remarks} 
\begin{enumerate}[\normalfont 1)]
   \item The explicit bounding constant of Theorem
  \ref{thm:classicalminimal velocity bound} is optimal. This may
  readily be
  seen by  examining  an almost-bounded orbit
  for the potential $V(x)=-C|x|^{-\mu}$, cf. \cite[Example
  2.2.4]{DeGe}, \cite{Ge} and
  \cite{Skib}.
\item The zero-energy orbits can behave somewhat unexpected like
  logarithmic spirals. (We encountered first such example in a preliminary
  version of the book \cite{DeGe}.)
\end{enumerate}
\end{remarks}

\subsection{Phase space localization}
\label{pseudodiff_Fefferman_Phong}
In this subsection we will prove \eqref{eq:PsDO_part}. This
is the main difficulty in proving Theorem
\ref{thm:resolvent_basic}.
The proofs of \eqref{eq:PsDO_part2} and \eqref{eq:disjoint_b}
are essentially identical to the present proof of
\eqref{eq:PsDO_part} and will be omitted.

Let us fix a real $\kappa'$ with
$\kappa < \kappa' < \kappa_0$, and consider the observable
$Q_s=\Op(q_s)$, where
\begin{align}
\label{eq:def_q}
q_s=q_s(x,\xi) = (w (\kappa'-b))^{s}\tilde{F}_{-}(b)
F_{-}(a_0); s\in {\mathbb R}.
\end{align}
We recall from \eqref{eq:definitions} that $w = \langle x \rangle f$; here and henceforth $f=f_{|\zeta|}$ with $\zeta \in \Gamma_{\theta}$ (as for $a_0$ and
$b$).
On $\supp (F_{-}(a_0))$ the symbol $b$
is bounded, and therefore $q_s \in S_{\it {unif}}\left(w^s,
g_{|\zeta|}\right)$.  We will prove the following result.

\begin{lemma}
\label{lem:Erik_Lem3}
For all $\epsilon > 0$, $s \geq 0$ the operator
$$
\langle x \rangle^{-1/2-\epsilon} f^{1/2} Q_s R(\zeta)
w^{-s} 
f^{1/2} \langle x \rangle^{-1/2-\epsilon} 
$$
is bounded uniformly in $\zeta \in \Gamma_{\theta}$.
\end{lemma}

\begin{proof}
Notice first that the statement for $s=0$ follows from
Corollary \ref{cor:unperturbed}. We will prove the lemma by
induction in $s$ using some ideas from the proof of
\cite[Lemma 2.6]{GIS}. 

Suppose we have proved Lemma \ref{lem:Erik_Lem3} for all
$s\leq s_0$ for some $s_0 \geq 0$. We will then prove that 
for all $\epsilon > 0$ and $s\in \left (s_0, s_0 + 1/2\right )$ the operator 
\begin{equation}
  \label{eq:1002}
 f Q_{s-1/2} R(\zeta) f w^{-s-1/2-\epsilon} \;\rm{is \; uniformly
  \; bounded}. 
\end{equation}
By the calculus of pseudodifferential operators and Corollary
\ref{cor:unperturbed} this is
equivalent to the statement of the lemma for $s$ in the same
range, cf. the proof of Lemma \ref{lem:auxiliary}. 

We may without loss of
generality assume that $F_{-}$ and $\tilde{F}_{-}$ are
real-valued, and (by the $\Psi$DO calculus) that 
\begin{align}
\label{eq:ass_supp_F}
&F_{-} =1 \text{ on a neighbourhood
of }[0, C_0], \\
\label{eq:ass_diff_F}
& \tilde{F}_{-} \text{ is decreasing and } \supp
(\tilde{F}_{-}') \subset (-\kappa, \kappa),
\end{align} 
where $C_0$ is the constant from Theorem \ref{thm:resolvent_basic} (\ref{item:C-0}).

We will place us in a situation, where we can apply the
Fefferman-Phong inequality (i.e. Theorem \ref{thm:CV_and_FPh_uniform} (\ref{Lowbound})) as
follows: Clearly $q_s^2 \in S_{\it {unif}}\left(
w^{2s}, g_{|\zeta|}\right)$, and therefore, since (by
\eqref{eq:symbolclass_h}) $h\in
S_{\it{unif}}(f^2\langle \frac {\xi}{f}\rangle^2,g_{|\zeta|})$ and 
$\langle \frac {\xi}{f}\rangle$ is bounded on ${\rm{supp}}(F_{-}(a_0))$,
$$
\{ h , q_{s}^2 \}_P \in S_{\it{unif}}\left(f^2
w^{2s-1},
g_{|\zeta|}\right).
$$
We will estimate the bracket  from above by a $\sigma \in
S_{\it {unif}}(f^2w^{2s-1},
g_{|\zeta|})$. We
then get as an operator inequality on $L^2({\mathbb R}^{d})$
$$
w^{-(s-3/2)} f^{-1}\Op(\sigma-\{ h , q_{s}^2 \}_P) 
f^{-1}w^{-(s-3/2)} \geq -C.
$$
The effective form of this estimate suited for implementing the
induction hypothesis will be: For any $m>0$
\begin{align}
  \label{eq:effect}
 \Op(\sigma-\{ h , q_{s}^2 \}_P) 
 \geq -CB^*B-C_m\langle x \rangle^{-m}, 
\end{align}
where $B=\langle x
\rangle^{-1/2} w^{s-s_o-1}f^{1/2}Q_{s_0}^1$ with $Q_{s_0}^1$ given as
the 
quantization of a symbol
$q_{s_0}^1$ of the form \eqref{eq:def_q} with the functions
${F}_{-}$ and $\tilde{F}_{-}$  replaced by say  
${F}_{-}^1$ and $\tilde{F}_{-}^1$; these functions  obey
\eqref{eq:ass_supp_F} and 
\eqref{eq:ass_diff_F} but they are `larger' than the previous ones.

Let us first calculate the principal symbol $\{h, q_{2s} \}_P$ of $i[H, Q_s^* Q_s]$. 
Here and henceforth we put, with a slight abuse of notation, $q_{2s} = q_s^2$. 
We decompose 
\begin{align}
\label{eq:Poisson_bracket}
\{h, q_{2s} \}_P  &= T_1+T_2+T_3;\\
T_1&=2
s \left(w(\kappa' -b)\right)^{2s-1} \tilde{F}_{-}^2(b)
F_{-}^2(a_0) \{h,  w (\kappa' -b)\}_P, \nonumber \\
T_2&= 2 \left(w (\kappa' -b)\right)^{2s} \tilde{F}_{-}(b)
\tilde{F}_{-}'(b) F_{-}^2(a_0) \{h, b\}_P, \nonumber \\
T_3&= 2 \left(w (\kappa' -b)\right)^{2s} \tilde{F}_{-}^2(b)
F_{-}(a_0)F_{-}'(a_0) \{h, a_0\}_P.\nonumber 
\end{align}

Defining $q_{2s-1} = (w
(\kappa'-b))^{2s-1}\tilde{F}_{-}^2(b) F_{-}^2(a_0)$ we get from the
computation in \eqref{eq:classgder}
\begin{align}
\label{eq:first_positive}
   T_1
= 2s q_{2s-1} ( 2 \kappa'v b -2h - W).
\end{align}
By \eqref{eq:positive_classic} 
the second term in \eqref{eq:Poisson_bracket} becomes, cf.  \eqref{postq}, 
\begin{align}
\label{eq:second_positive}
T_2
=2 \hat{q}_{2s-1}
\tilde{F}_{-}'(b) (2h + W -2b^2v),
\end{align}
where $\hat{q}_{2s-1} = w^{-1} \left(w(\kappa' -b)\right)^{2s}
\tilde{F}_{-}(b) F_{-}^2(a_0)$.

Finally for the third term in
\eqref{eq:Poisson_bracket} we may write with some symbol
$\check{q}_{2s-1} \in S_{\it {unif}}\left(w^{2s-1}, g_{|\zeta|}\right)$ 
\begin{align}
\label{eq:third_positive}
T_3
=
f^2\check{q}_{2s-1} F_{-}'(a_0).
\end{align}

The rest of the calculation is split in two depending on the relative
size of $\Re \zeta$ and $|\zeta|$.

{\bf Case 1. $ \Re \zeta \geq \frac{(\kappa')^2}{\kappa_0^2}
|\zeta|$.} Using this, the fact that $b < \kappa$ on $\supp
(q_{2s-1})$ and \eqref{eq:estim_W} we may estimate the
right hand side of \eqref{eq:first_positive}
\begin{align}
\label{eq:Pos_comm_pseudo}
&2s q_{2s-1} ( 2 \kappa'v b -2h - W) \nonumber \\
& \leq
2s q_{2s-1} \left\{2 \kappa \kappa'(|\zeta|/\kappa_0^2 +
\langle x
\rangle^{-\mu}) - 2 \Re \zeta - 2 \Re(h-\zeta) - 2 \kappa_0^2 \langle x
\rangle^{-\mu}\right\} \nonumber \\
&\leq -\delta q_{s-1/2} f^2 q_{s-1/2}
- 2s \left\{ (h-\overline{\zeta}) q_{2s-1} +
q_{2s-1}(h-\zeta)\right\},
\end{align}
here with $\delta=\delta(\kappa, \kappa',\kappa_0, s, \mu) >0$ and 
$$q_{s-1/2}(x,\xi) = (w
(\kappa'-b))^{s-1/2}\tilde{F}_{-}(b) F_{-}(a_0).$$
Since $f^{-2}h$ is bounded on $\supp (q_{2s-1})$ the right
hand side of \eqref{eq:Pos_comm_pseudo} clearly lies in 
$S_{\it{unif}}\left(f^2 w^{2s-1}, g_{|\zeta|}\right)$.

To estimate the right hand side of
\eqref{eq:second_positive} we use the property
\eqref{eq:ass_diff_F}. Thus $\tilde{F}_{-}' \leq 0$ and we
have $b^2 \leq \kappa^2$ on $\supp (\tilde{F}_{-}'(b))$. So we
see as above that
\begin{align}
\label{eq:Pos_comm_pseudo2}
2 \hat{q}_{2s-1}
\tilde{F}_{-}'(b) (2h + W -2b^2v) 
\leq
4 \hat{q}_{2s-1} \tilde{F}_{-}'(b) \Re(h-\zeta).
\end{align}
Clearly the right hand side of
\eqref{eq:Pos_comm_pseudo2} is a symbol in 
$S_{\it{unif}}\left(f^2 w^{2s-1},
g_{|\zeta|}\right)$. 

The input to our application of the Fefferman-Phong
inequality is therefore the estimate (combining
\eqref{eq:Poisson_bracket} -
\eqref{eq:Pos_comm_pseudo2}) 
\begin{align}
\label{101}
&\delta (f q_{s-1/2})^2
\leq  
-\{ h , q_{2s} \}_P - 2s \left\{
(h-\overline{\zeta}) q_{2s-1} + q_{2s-1}(h-\zeta)\right\}\nonumber \\
&+
2\left\{
(h-\overline{\zeta})
\hat{q}_{2s-1} \tilde{F}_{-}'(b) +\hat{q}_{2s-1}
\tilde{F}_{-}'(b) (h-\zeta)\right\} 
+ f^2\check{q}_{2s-1} F_{-}'(a_0) \nonumber \\
&=
-\{ h , q_{2s} \}_P\\
 &\;\;\;\;\;\;\;\;\;\;\;\;\;\;\;+\left\{
(h-\overline{\zeta})
q_{2s-1}^{{\rm final}} 
+q_{2s-1}^{{\rm final}}(h-\zeta)\right\} +f^2\check{q}_{2s-1} F_{-}'(a_0),\nonumber 
\end{align}
with
$$
q_{2s-1}^{{\rm final}}  =
2\hat{q}_{2s-1}\tilde{F}_{-}'(b) - 2sq_{2s-1}.
$$

To show \eqref{eq:1002} we apply \eqref{eq:effect} and  \eqref{101}. We introduce $R=R(\zeta)$, $T^{-1} = f w^{-s-1/2-\epsilon}$,
$\phi\in L^2$, $\|\phi\|=1$  and $\psi=RT^{-1} \phi$, and use the
induction hypothesis and Corollary \ref{cor:unperturbed} to  obtain
\begin{align}
\label{eq:pos_commutator}
  \delta 
\| f Q_{s-1/2} \psi \|^2 -
C_1&\leq 
  \delta \left\langle \Op((q_{s-1/2} f)^2)
  \right\rangle_{\psi}  \nonumber \\
  &\leq 
  -\left\langle \Op(\{ h , q_{2s} \}_P)
  \right\rangle_{\psi} \\
  &+ 2 \left\langle \Op\left( (h-\overline{\zeta})
q_{2s-1}^{{\rm final}} + q_{2s-1}^{{\rm
final}}(h-\zeta)\right)
  \right\rangle_{\psi} \nonumber \\
  &
+
\left\langle \Op(f^2\check{q}_{2s-1} F_{-}'(a_0))
  \right\rangle_{\psi}+ C_2. \nonumber
\end{align}
On the right hand side of \eqref{eq:pos_commutator} we consider each
term separately. For the first term we know that $\{ h ,
q_{2s} \}_P$ is the
principal symbol of the pseudodifferential operator 
$i [ H  , Q_{2s} ]$; $Q_{2s}= Q_{s}^* Q_{s}$.
Thus, the pseudodifferential calculus and the induction
hypothesis yield
\begin{align}
\label{eq:FeffermanPhong}
   - \langle \Op(\{ h , q_{2s} \}_P) \rangle_{\psi}\leq -\langle
  i[H,Q_{2s}]\rangle_{\psi} +C_3.
\end{align}
Clearly
$$-\langle
  i[H,Q_{2s}]\rangle_{\psi} =2 \Im \langle T^{-1} \phi, Q_{2s} \psi \rangle -
  2 \Im(\zeta) \|Q_s \psi \|^2.$$ 
Since $\Im \zeta > 0$ we may drop the second term. Thus we  conclude from
\eqref{eq:FeffermanPhong}  (after an application of the
pseudodifferential calculus, the induction
hypothesis  and 
Cauchy-Schwarz) that 
\begin{align}
\label{eq:Cauchy-Schwarz}
   &- \langle \Op(\{ h , q_{2s} \}_P) \rangle_{\psi}\nonumber\\
  &\leq 
  \eta \|f Q_{s-1/2} \psi\|^2 +
  \eta^{-1} \|f^{-1} Q_{s+1/2} T^{-1} \phi
  \|^2 + C_4;\;\eta = \delta/2.
\end{align}
Notice that with
$T^{-1} = f w^{-s-1/2-\epsilon}$ the second term on the right
hand side is clearly bounded since it does not contain a
resolvent.

The second term in
\eqref{eq:pos_commutator} is similar but easier. We
calculate as above,
\begin{align}
\label{eq:Cauchy-Schwarz2}
&  \left\langle \Op\left( (h-\overline{\zeta})
q_{2s-1}^{{\rm final}} + 
q_{2s-1}^{{\rm final}}(h-\zeta)\right)
\right\rangle_{\psi} \\
&\leq 
2 \Re \langle  (H-\zeta) \psi ,
\Op(q_{2s-1}^{{\rm final}}) \psi \rangle +
C_5\nonumber \\
&\leq \| f^{-1} w^{s+1/2 + \epsilon}  T^{-1} \phi \| 
\times \| f w^{-s-1/2 - \epsilon} \Op(q_{2s-1}^{{\rm
final}}) \psi \|
 + C_5. \nonumber
\end{align}
The final expression can clearly be estimated by  using the
induction hypothesis. 

The third  term in \eqref{eq:pos_commutator} is
easily seen to be bounded using the property  
\eqref{eq:ass_supp_F} and Lemma \ref{lem:auxiliary}. Therefore, inserting
\eqref{eq:Cauchy-Schwarz} and
\eqref{eq:Cauchy-Schwarz2} in
\eqref{eq:pos_commutator} proves boundedness of $\|f
Q_{s-1/2} \psi
\|$, which is what we aimed at.

{\bf Case 2. $ \Re \zeta < \frac{(\kappa')^2}{\kappa_0^2}
|\zeta|$.}
In this case $2 \Re \zeta - \frac{2\kappa \kappa'}{\kappa_0^2}
|\zeta|$ can be negative. So instead of writing $2h + W = 2
\Re(h-\zeta) + (2\Re \zeta + W)$ as in \eqref{eq:Pos_comm_pseudo}, we
write
$$
2h + W = \left[ 2 \Re(h-\zeta) + C \Im(h-\zeta) \right] + (2 \Re \zeta
+ C \Im \zeta + W),
$$
for $C > 0$. Using that $\Re \zeta < \frac{(\kappa')^2}{\kappa_0^2}
|\zeta|$, we choose $C$ so big that
$$
2 \Re \zeta + C \Im \zeta - \frac{2\kappa \kappa'}{\kappa_0^2}
|\zeta| \geq \Im \zeta \geq \delta'| \zeta|.
$$
Thus, instead of \eqref{eq:Pos_comm_pseudo}, we find, for some
$\delta''>0$,
\begin{align*}
&2s q_{2s-1} ( 2 \kappa'v b -2h - W)  \\
&\leq -\delta'' q_{s-1/2} f^2 q_{s-1/2}
- 2s \left\{ (h-\overline{\zeta}) q_{2s-1} +
q_{2s-1}(h-\zeta)\right\} \\
&
-iCs \left\{ (h-\overline{\zeta}) q_{2s-1} -
q_{2s-1}(h-\zeta)\right\}.
\end{align*}
The same thing is done in \eqref{eq:Pos_comm_pseudo2}.
The rest of the proof is now similar to Case 1.

\end{proof}

We now remove the extra $b$'s appearing in the statement of Lemma
\ref{lem:Erik_Lem3} compared to \eqref{eq:PsDO_part}. Notice that this follows readily from the pseudodifferential
calculus  since $(\kappa'-b)^{-s}$ is bounded on $\supp
(F_{-}(b) F_{-}(a_0))$ (this was in fact also used in the proof of Lemma
\ref{lem:Erik_Lem3}).

\begin{lemma}
\label{lem:Erik_Lem4}
Let $s \geq 0$ be arbitrary. Let $A = A_\zeta=
\Op(a)$ be the quantization  of the symbol
$a=  \tilde F_{-}(b) F_{-}(a_0)$. Then for all
$\epsilon > 0$ the operator
$$
\langle x \rangle^{-\epsilon} f w^{s-1/2} A
R(\zeta) w^{-s-1/2} f \langle x \rangle^{-\epsilon},
$$
is bounded uniformly in $\zeta \in \Gamma_{\theta}$.
\end{lemma}

We can now finish the proof of \eqref{eq:PsDO_part}:

\begin{proof}[Proof of \eqref{eq:PsDO_part}]
The final step in proving \eqref{eq:PsDO_part} consists of
replacing the weights $f$ and $ w$ in Lemma \ref{lem:Erik_Lem4} by their
limits as $|\zeta|$ goes to zero. That is the content of Lemma
\ref{lem:changing_weights}.  
\end{proof}

\section{Proof of main results}
\label{iterated_perturbed}

\subsection{Proof of Theorem \ref{thm:lap}} 
\label{snik-snak}
We
will only consider the case $\zeta \in
\Gamma_{\theta}$---the other case, $\overline{\zeta} \in
\Gamma_{\theta}$, can be proved analogously.

Let us write for $\zeta \in {\mathbb C}\setminus
{\mathbb R}$
\begin{align}
\label{eq:5}
  R(\zeta) = (H-\zeta)^{-1},\; R_1(\zeta) = (H_1-\zeta)^{-1},
\end{align}
with $H_1 = p^2 + V_1 = H - V_2$.
We shall proceed perturbatively using
\begin{equation}
   \label{eq:6}
   R(\zeta) (I + V_2 R_1(\zeta)) = R_1(\zeta).
\end{equation}
First we show that $R_1(\zeta)$ is uniformly H{\"o}lder continuous in
$\Gamma_{\theta}$.
For that we interpolate \eqref{eq:x_weights_iterated} for $m=1$ and
$m=2$.
We consider the family of bounded operators
\begin{align}
  \label{eq:x_weights_iterateddif}
B(z)=k^{-z-\epsilon} \{R_1(\zeta_1)-R_1(\zeta_2)\}
k^{-z-\epsilon};\;\Re(z)\in [1/2,3/2].
\end{align}
For 
$\Re(z)=3/2$ (using $R_1(\zeta_1)-R_1(\zeta_2) = \int_0^1 \frac{d}{dt}R_1(\zeta_2+t(\zeta_1-\zeta_2))\,dt$)
we have the bound $\|B(z)\|\leq C {|}\zeta_1-\zeta_2|$,
and for $\Re(z)=1/2$ the bound $\|B(z)\|\leq C$, yielding 
\begin{align}
  \label{eq:x_weights_iterateddif2}
\|B(z)\|\leq C|\zeta_1-\zeta_2|^{\Re (z)-1/2}.
\end{align}
For $s\leq 3/2(1+\mu/2)$ we choose $z=\Re (z)=s(1+\mu{/2})^{-1}$ in
\eqref{eq:x_weights_iterateddif2}; otherwise  we take $z=3/2$.
This proves the H{\"o}lder continuity statement of Theorem
\ref{thm:lap}
in the case $V_2=0$. In particular $R_1^+ = R_1(0+i0)=\lim_{\zeta\rightarrow 0, \zeta
\in \Gamma_{\theta}}R_1(\zeta)$ and $R_1^- = R_1(0-i0)=\lim_{\zeta\rightarrow 0, \zeta
\in \Gamma_{\theta}}R_1\!\!\left(\overline{\zeta}\right)$ are
well-defined
(in weighted spaces).

To show \eqref{eq:limitbound} (in the general case) it suffices to show that
$\langle x
\rangle^{s} ( I + V_2 R_1^{+})
\langle x \rangle^{-s}$ is invertible as an operator on
$L^2({\mathbb R}^d)$. (Here we use \eqref{eq:6}, as
well
 as the standard limiting absorption principle for
positive
energies and absence of positive eigenvalues, cf. \cite{mourre}, \cite{Tam}
and \cite[Section 6.5]{DeGe}.)

Notice that $\langle x \rangle^{s} V_2
R_1^{+}\langle x
\rangle^{-s}$ is compact (being the norm-limit of a compact operator-valued
function). 
 Whence by Fredholm theory, it suffices to show that the equation
\begin{align}
   \label{eq:fredholm}
\phi = - V_2 R_1^+ \phi,
\end{align}
has no nonzero solution
$\phi \in \langle x \rangle^{-s} L^2({\mathbb R}^d)$.
Let
$\psi = R_1^+
\phi \;(\in \langle x \rangle^{s} L^2({\mathbb R}^d))$. Then we have  in the sense of distributions
\begin{align}
\label{eq:psi_loeser_en_ligning}
   H \psi = 0 \;\rm{and}\;V_2 \psi = -\phi.
\end{align}

We can calculate
\begin{align*}
0&= \Im \langle \psi, V_2 \psi \rangle =
- \Im \langle \psi, \phi \rangle  \\
&= - \Im \langle R_1^+ \phi, \phi \rangle
= (2i)^{-1}\langle \phi, (R_1^+- R_1^{-})\phi \rangle.
\end{align*}
Since $\frac{R_1^+- R_1^{-}}{2i} \geq 0$ we get that
\begin{eqnarray}
  \label{eq:ident}
  \psi = R_1^{+} \phi = R_1^{-} \phi.
\end{eqnarray}

\begin{lemma}
  \label{lem:agmon} $\psi\in L^2({\mathbb R}^d)$.
\end{lemma}
\begin{proof}
 A priori $\phi\in \langle x \rangle^{-s}L^2$ and $\psi\in
\langle x \rangle^{s^{\prime}}L^2;\;s^{\prime}>\frac
{1}{2}+\frac {\mu}{4}$. From \eqref{eq:psi_loeser_en_ligning}
we learn that if $\psi\in\langle x \rangle^{s^{\prime}}L^2$
for some real $s^{\prime}$ then  $\phi\in\langle x
\rangle^{s^{\prime}-1-\mu/2-\delta}L^2$. In particular we
have  $\phi\in \langle x \rangle^{-\tilde s}L^2;\;\tilde
s<\frac {1}{2}+\frac {\mu}{4}+\delta$.  The idea of the
proof is to show by a bootstrap argument that we may take
$s^{\prime}$ arbitrary. A bootstrap argument for a similar
problem for the free Laplacian was given by Agmon in his
proof of \cite[Theorem 3.3]{Ag}. Our analysis is based on
Theorem \ref{thm:resolvent_basic}.

We pick a real-valued function  ${F}_{+}$ as in Theorem
\ref{thm:resolvent_basic} (\ref{item:C-0}) such that $F_+(x)=1$ for
$|x|>2C_0$. Let ${F}_{-}=1-{F}_{+}$. Pick real-valued
functions 
$\tilde{F}_{-}$ and $\tilde{F}_{+}$ as in
Theorem \ref{thm:resolvent_basic} (\ref{some_label}) such that 
$\tilde{F}_{-}+\tilde{F}_{+}=1$. Then we decompose with the
symbols $a_0$ and $b$ being defined as in
\eqref{eq:def_a0_b} with $E=0$ in the expression
\eqref{eq:def_f_2} for $f$
\begin{align}
  \label{eq:decom}
 \psi& =\Op(
 {F}_{+}(a_0))\psi+\Op({F}_{-}(a_0)\tilde{F}_{-}(b))\psi\nonumber\\& +
\Op({F}_{-}(a_0)\tilde{F}_{+}(b))\psi.
\end{align}
By \eqref{eq:quantum_part} and \eqref{eq:PsDO_part} the
first two terms on the right hand side of \eqref{eq:decom}
belong to $\langle x \rangle^{s^{\prime}}L^2$ where (assuming here $\phi\in \langle x \rangle^{-s}L^2$)
\begin{eqnarray}
  \label{eq:connection}
s^{\prime}=(1+\tfrac{\mu}{2})(-t+\tfrac{1}{2}+\epsilon);\;t=\frac{s}{1+\tfrac{\mu}{2}}-\tfrac{1}{2}-\epsilon.  
\end{eqnarray}
We
notice that 
the bound \eqref{eq:PsDO_part2} is equivalent to 
\begin{align}
\label{eq:PsDO_part22}
&\|k^{t-1/2-\epsilon} \Op(F_{-}(a_0) \tilde{F}_{+}(b))R_1(\zeta)^*
k^{-t
  -1/2-\epsilon}\| \leq C.
\end{align}
Taking $\zeta\to 0$ in the sector $\Gamma_\theta$, \eqref{eq:PsDO_part22}
leads to 
\begin{align}
\label{eq:PsDO_part222}
&\|k^{t-1/2-\epsilon} \Op(F_{-}(a_0) \tilde{F}_{+}(b))R_1^{-}
k^{-t
  -1/2-\epsilon}\| \leq C,
\end{align}
with the same convention for  $a_0$ and $b$ as above.
We use the representation $\psi = R_1^{-} \phi$ of \eqref{eq:ident}
and apply \eqref {eq:PsDO_part222},
 and conclude that also the third term
on the right hand side of \eqref{eq:decom} belongs to
$\langle x \rangle^{s^{\prime}}L^2$ with $s^{\prime}$ given by
\eqref{eq:connection}; so $\psi \in \langle x \rangle^{s^{\prime}}L^2$. 

From this we learn that
$$\phi\in\langle x \rangle^{s^{\prime}-1-\frac{\mu}{2}-\delta}L^2=\langle x
\rangle^{-s-\delta+(2+\mu)\epsilon}L^2;$$
so by taking $\epsilon<(2+\mu)^{-1}\delta$ we  improve the
decay of $\phi{}$. 
 Iterating
this argument (gaining at each iteration almost a factor
$\langle x \rangle^{-\delta{}}$) leads to $s^{\prime}\leq 0$ eventually. 
\end{proof}

Combining Theorem \ref{thm:absence} and Lemma \ref{lem:agmon} yields
$\psi=\phi=0$,
completing the proof of \eqref{eq:limitbound} in the general case.
It remains to show the H{\"o}lder continuity statement of Theorem
\ref{thm:lap} in the general case. This may easily be done using
\eqref{eq:6} and the known result for $R_1(\zeta)$; we omit the details.

\begin{remark}
\label{rem:virialtypeargument}
There exists another approach to proving \eqref{eq:limitbound} based
on a virial type argument and Theorem \ref{thm:absence}: Formally, on one hand 
$$\langle\psi,i[H_1,A]\psi\rangle= \langle\psi,W\psi\rangle,$$
while on the other hand
$$\langle\psi,i[H_1,A]\psi\rangle= -2\Im (\langle\phi,A\psi\rangle).$$ 
This leads to the conclusion (rigorously, after some work using again \eqref{eq:ident}) that 
$p_j \psi, \langle x \rangle^{-\mu/2} \psi \in L^2({\mathbb
R}^d)$. Therefore, we can apply Theorem \ref{thm:absence} to
conclude that $\psi$, and therefore $\phi$, vanish
identically. To make this work one needs a stronger decay assumption than Theorem
\ref{thm:lap} (\ref{it:assumption4}).  As an advantage, being
independent of Theorem \ref{thm:resolvent_basic}, 
this
method only requires a few derivatives of $V_1$, cf. Remark
\ref{rem:hoelder}.
\end{remark}

\subsection{Proof of Theorem \ref{thm:lap_iterated}} 
\label{snik-snak2}
We will only prove \eqref{eq:iterated_estimate}, since the
H\"{o}lder continuity follows from
\eqref{eq:iterated_estimate} using interpolation as in
Subsection \ref{snik-snak}.

Let $V = V_1 + V_2$ be the
decomposition of $V$ given in the statement of Theorem
\ref{thm:lap_iterated}.
Two successive applications of the resolvent identity \eqref{eq:6} give
\begin{align}
\label{eq:resolventidentity_twice}
  R(\zeta) = R_1(\zeta) - R_1(\zeta) V_2 R_1(\zeta) + 
  R_1(\zeta) V_2 R(\zeta) V_2 R_1(\zeta).
\end{align}
We will prove by induction in $m$ that
\begin{align}
\label{eq:induction_lap_iterated}
  \left\| k^{-(m-1/2)-\epsilon} R_1(\zeta)^s R(\zeta)^t
  k^{-(m-1/2)-\epsilon} \right\| \leq C,
\end{align}
for $s,t \in {\mathbb N}\cup \{0\}$ with $1 \leq s+t \leq m$.
It is clear from Theorem \ref{thm:lap} that
\eqref{eq:induction_lap_iterated} holds for $m=1$. So let
us assume that
\eqref{eq:induction_lap_iterated} holds for all $m \leq m_0$ and prove
that it holds for $m=m_0+1$. 

The case $s=m_0+1$, $t=0$ is statement
\eqref{eq:x_weights_iterated} of Theorem
\ref{thm:iterated_resolvents}. So suppose
\begin{align}
\label{eq:induction_theorem1.4}
\|k^{-(m_0+1/2)-\epsilon} R_1(\zeta)^s R(\zeta)^t
  k^{-(m_0+1/2)-\epsilon}\|\leq C,
\end{align}
where $s \geq \sigma$ and $s+t =
m_0 +1$.
Then need to bound
\begin{align}
\label{eq:induction_induction}
k^{-(m_0+1/2)-\epsilon} R_1(\zeta)^{\sigma - 1}
R(\zeta)^{\tau+1}
  k^{-(m_0+1/2)-\epsilon};\;\tau = m_0 - \sigma + 1.
\end{align}

Upon substitution of
\eqref{eq:resolventidentity_twice} the expression
\eqref{eq:induction_induction} becomes
\begin{align*}
&k^{-(m_0+1/2)-\epsilon} R_1(\zeta)^{\sigma} 
R(\zeta)^{\tau}
  k^{-(m_0+1/2)-\epsilon} \\
- &k^{-(m_0+1/2)-\epsilon} R_1(\zeta)^{\sigma} V_2 R_1(\zeta)
R(\zeta)^{\tau}
  k^{-(m_0+1/2)-\epsilon} \\
+&
k^{-(m_0+1/2)-\epsilon} R_1(\zeta)^{\sigma} V_2
R(\zeta) V_2 R_1(\zeta) R(\zeta)^{\tau}
  k^{-(m_0+1/2)-\epsilon} \\
=& E_1 + E_2 + E_3.
\end{align*}

By the hypothesis \eqref{eq:induction_theorem1.4}, $E_1$ is
 uniformly  bounded.

To estimate $E_2$ we write (with
$N$ sufficiently big)
\begin{align*}
E_2 &=
\left( k^{-(m_0+1/2)-\epsilon} R_1(\zeta)^{\sigma} \langle
x \rangle^{-N} \right)
\left( \langle x \rangle^{N} V_2 R_1(\zeta)
R(\zeta)^{\tau} k^{-(m_0+1/2)-\epsilon} \right).
\end{align*}
We only need to estimate the last factor.
By using the resolvent equation it may written as 
\begin{align*}
&BB_1+(\zeta-i)BB_2;\;B=\langle x \rangle^{N} V_2 R_1(i) \langle x
\rangle^{N},\\
&B_1=\langle x \rangle^{-N}
R(\zeta)^{\tau} k^{-(m_0+1/2)-\epsilon},\;B_2=\langle x \rangle^{-N} R_1(\zeta)
R(\zeta)^{\tau} k^{-(m_0+1/2)-\epsilon}.
\end{align*}
Using the relative
boundedness of $V_2$ and the property of compact support, we see
that $B$  is bounded. The two other
factors $B_1$ and $B_2$  are bounded by the induction hypotheses.

The argument for the term $E_3$ is similar; it is omitted.

Thus we have a uniform bound of 
\eqref{eq:induction_induction}, and  \eqref{eq:induction_lap_iterated} follows.
Clearly the bound  \eqref{eq:induction_lap_iterated} with
$s=0$, $m = t$  and \eqref{eq:iterated_estimate} coincide, so the proof is finished. 

\subsection{Proof of  \eqref{eq:E ne 0}}\label{snakk}
Suppose on the contrary that  $E^{\prime}(+0)= 0$. Let $R^+=R(0+i0)$
and
$R^-=R(0-i0)$, so that $R^+=R^-$  as an identity in ${\mathcal B}({\mathcal
  H}_1, {\mathcal H}_2)$. We also use the notation $R_1^+$ and $R_1^-$ of
the proof of Theorem \ref{thm:lap} in
  Subsection \ref{snik-snak}.

We compute
  \begin{align*}
  &(2i)^{-1}\langle \phi, (R_1^+- R_1^{-})\phi \rangle \\
&=\lim_{\kappa\to 0^+}\kappa\| R_1(i\kappa)\phi\|^2\\
&=\lim_{\kappa\to 0^+}\kappa\| R(i\kappa)\tilde\phi(\kappa)\|^2;\;\tilde\phi(\kappa)=(I+V_2R_1(i\kappa))\phi,\\
&=(2i)^{-1}\langle  \tilde\phi(+0), (R^+-
R^{-})\tilde\phi(+0)\rangle =0, 
  \end{align*}
from which we conclude that also
\begin{equation}
  \label{eq:R_1=R_1}
  R_1^+= R_1^{-}.
\end{equation}

Using the proof of Theorem \ref{thm:lap} we may infer from
\eqref{eq:R_1=R_1} that for all $t>0$
\begin{equation}
  \label{eq:weightest}
k^{t -1/2-\epsilon}R_1^+k^{-t-1/2-\epsilon}\in{\mathcal B}(L^2({\mathbb R}^d)).
\end{equation}

Let us consider  vectors $\phi\in
L^2(B_\rho)$ where $B_\rho=\{x\in{\mathbb
    R}^d|\;|x|\leq \rho
\}$ with 
$\rho$ large. By taking $t =1/2+\epsilon$ in \eqref{eq:weightest} we
learn that $\psi:=R_1^+\phi\in L^2({\mathbb R}^d)$. Since $H_1\psi=\phi$ we see
that  $H_1\psi=0$ outside $B_\rho$. Using this and Remarks \ref{h}
\ref{it:Rresult}) we conclude that indeed $\psi=0$ outside
$B_\rho$.
 We conclude that $R_1^+\in{\mathcal B}(L^2(B_\rho))$ is the
 inverse of  the Dirichlet Laplacian $H_{1,\rho}=-\Delta +V_1$ for the
 region $B_\rho$.
We obtain a contradiction from this by choosing a large $\rho$ such that zero
is an eigenvalue of $H_{1,\rho}$. Notice that all branches of
eigenvalues will cross zero as $\rho\to\infty$, cf. \cite[Theorem XIII.6
(Vol. IV, p. 87)]{ReSi}.

\subsection{Proof of Theorem \ref{thm:minimal velocity}}
\label{snik-snak3}

$\vspace{0.1 cm}$

{\bf Proof of  (\ref{it:local decay estimate}).} 
Due to Theorem \ref{thm:lap_iterated} we may define
\begin{align*}
 E^{(m)}(\lambda)=\frac {d^{m-1}}{d\lambda^{m-1}}E^{\prime}(\lambda)&=(2\pi i)^{-1} \frac {d^{m-1}}{d\lambda^{m-1}}
\left(R(\lambda+i0)-R(\lambda-i0)\right)\\&\in {\mathcal B}(k^{-(m-1/2) -\tilde\epsilon} L^2,k^{(m-1/2) +\tilde\epsilon} L^2),
\end{align*}
and
$$E^{(m)}(+0)=\lim_{\lambda\to
  0^+}E^{(m)}(\lambda).$$

Using  similar notation for $(fE^{\prime})(\lambda)$ we represent as an operator in this space, cf. \cite{JMP},
\begin{align}
\label{eq:jensen}
&e^{-itH}(f1_{[0,\infty)})(H)=\int_0^{\infty}e^{-it\lambda}f(\lambda)E^{\prime}(\lambda)d\lambda{}
\\
&=\sum_{n=1}^{m-1}(it)^{-n}(fE^{\prime})^{(n-1)}(+0)+\int_0^{\infty}e^{-it\lambda}(it)^{-m+1}(fE^{\prime})^{(m-1)}(\lambda)d\lambda{}.
\nonumber 
\end{align}
 
Take $m=3$. 

{\bf Proof of  (\ref{it:minimal velocity bound}).}
We multiply \eqref{eq:jensen} by $F(\langle x\rangle <2t^\kappa)$ (where
$\kappa=(1-\epsilon)(1+\mu /2)^{-1}$) from the left and
$k^{-(m-1/2) -\tilde\epsilon}$ from the right. Using again  
Theorem
  \ref{thm:lap_iterated} 
we estimate 
\begin{align*}
 &\|F(\langle x \rangle<2t^\kappa)\int_0^{\infty}e^{-it\lambda}(it)^{-m+1}(fE^{\prime})^{(m-1)}(\lambda)d\lambda{}k^{-(m-1/2) -\tilde\epsilon}\| \\
&\leq C_1\int_0^{\infty}\|(\frac {t^{1-\epsilon}}{k})^{(m-1/2)
  +\tilde\epsilon}
  e^{-it\lambda}(it)^{-m+1}(fE^{\prime})^{(m-1)}(\lambda)k^{-(m-1/2)
  -\tilde\epsilon}\|d\lambda{}\\
&\leq C_2t^{\frac {1}{2}+\tilde\epsilon(1-\epsilon) -\epsilon(m-\frac {1}{2})}.
\end{align*}

Suppose
$$\frac {1}{2}-\epsilon(m-\frac {1}{2})<-(1+\epsilon^{\prime})\frac {1}{2}.$$
Then for all sufficiently small $\tilde\epsilon>0$

$$\frac {1}{2}+\tilde\epsilon(1-\epsilon) -\epsilon(m-\frac
{1}{2})\leq -(1+\epsilon^{\prime})\frac {1}{2}.$$
This is the argument for the contribution from the last term on the
right hand side of \eqref{eq:jensen}.

We deal with the other terms in
a similar way:
\begin{align*}
 &\|F(\langle x \rangle<2t^\kappa)(it)^{-n}(fE^{\prime})^{(n-1)}(+0)k^{-(m-1/2) -\tilde\epsilon}\| \\
&\leq C_1\|(\frac {t^{1-\epsilon}}{k})^{(n-1/2)
  +\tilde\epsilon}
  (it)^{-n}(fE^{\prime})^{(n-1)}(+0)k^{-(m-1/2)
  -\tilde\epsilon}\|\\
&=C_2t^{(1-\epsilon)((n-1/2) +\tilde\epsilon)}t^{-n}
\\
&=C_2t^{ \tilde\epsilon(1-\epsilon)+(\frac {1}{2}-n)\epsilon-\frac {1}{2}}.
\end{align*}
The worst bound is for $n=1$. Clearly we get the result.

\appendix

\section{Algebraic verification of Theorem
\ref{thm:iterated_resolvents}}
\label{algebra}

We prove the equations \eqref{group2} by induction in $m$.
For $m=1$ the equations \eqref{group2} become
\eqref{group1}. So assume that
\eqref{group2} are true for all $m \leq m_0$ for some $m_0
\geq 1$. We will prove \eqref{eq:x_weights_iterated},
\eqref{eq:quantum_part_iterated},
\eqref{eq:PsDO_part_iterated},
\eqref{eq:disjoint_b_iterated},
and \eqref{eq:disjoint_a_iterated} for
$m=m_0+1$. By symmetry (look at the `adjoint' expressions)
\eqref{eq:quantum_part222_iterated},
\eqref{eq:PsDO_part2_iterated} and
\eqref{eq:disjoint_a222_iterated} will follow from
the previous proofs, so these cases will not need further
elaboration.
For shortness we write
$R=R(\zeta)$.

{\bf Proof of \eqref{eq:x_weights_iterated}}.
To prove \eqref{eq:x_weights_iterated} for $m=m_0+1$ we
choose a real-valued function  ${F}_{+}$ as in Theorem
\ref{thm:resolvent_basic} (\ref{item:C-0}) such that $F_+\equiv 1$ on
a neighbourhood of $+\infty$. Let $F_{-}=1-{F}_{+}$. Pick
real-valued functions
$\tilde{F}_{-}$ and $\tilde{F}_{+}$ as in
Theorem \ref{thm:resolvent_basic} (\ref{some_label}) such that
$\tilde{F}_{-}+\tilde{F}_{+}=1$. Then obviously $1=
F_{+}(a_0) + F_{-}(a_0)\tilde{F}_{+}(b) +
F_{-}(a_0)\tilde{F}_{-}(b)$.
 We decompose
\begin{align}
\label{eq:split}
  & k^{-(m_0+1/2)-\epsilon} R^{m_0+1} k^{-(m_0+1/2)-\epsilon}\nonumber
\\&=
\left(k^{-(m_0+1/2)-\epsilon} Rk^{-s}\right)
\left(k^{s}
\Op(F_+(a_0)) R^{m_0}   k^{-(m_0+1/2)-\epsilon}\right)
\\
&+
\left(k^{-(m_0+1/2)-\epsilon} R k^{-s}\right)
\left(k^{s} \Op(F_{-}(a_0)\tilde{F}_{-}(b)) R^{m_0}
k^{-(m_0+1/2)-\epsilon}\right) \nonumber \\
&+ \left(k^{-(m_0+1/2)-\epsilon} R
  \Op(F_{-}(a_0)\tilde{F}_{+}(b))k^s \right)
\left( k^{-s} R^{m_0}
k^{-(m_0+1/2)-\epsilon}\right),\nonumber
\end{align}
 cf. \eqref{eq:decom}.
Choosing
$s=1/2+\epsilon/3$, the first term in \eqref{eq:split}
is seen to be uniformly bounded by using
\eqref{eq:x_weights} and \eqref{eq:quantum_part_iterated} with $m=m_0$
and $t=1+\epsilon2/3$ and
 $\epsilon \to \epsilon/3$.
 The second term is treated
similarly using \eqref{eq:PsDO_part_iterated} instead of
\eqref{eq:quantum_part_iterated}.
For the third term we choose $s=m_0+\epsilon/3-1/2$
and
apply \eqref{eq:PsDO_part2} and \eqref{eq:x_weights_iterated} to get the
conclusion.

In the rest of the proof we will not explicitly introduce
 factors $k^s$ and $ k^{-s}$ as we did above.

{\bf Proof of \eqref{eq:quantum_part_iterated}}.
For this part we introduce functions
$G_{+}, \tilde{G}_{+},\tilde{G}_{-}$ analogous to the $F$'s
in the argument above and satisfying the same conditions. We
assume that
$G_{+}
\equiv 1$ on a neighbourhood of $\supp F_{+}$. Let $G_{-} =
1 - G_{+}$. Then we write with $B= \Op(F_{+}(a_0))$ and $\tau
= t+m_0+1/2+\epsilon$:
\begin{align}
\label{eq:split2}
  k^{t-1/2-\epsilon} BR^{m_0+1}
k^{-\tau} 
=&
k^{t-1/2-\epsilon} BR \Op(G_{+}(a_0)) R^{m_0}
k^{-\tau} \\
&+ k^{t-1/2-\epsilon} BR
\Op(G_{-}(a_0)\tilde{G}_{-}(b))  R^{m_0}
k^{-\tau} \nonumber\\
&+
k^{t-1/2-\epsilon} BR
\Op(G_{-}(a_0)\tilde{G}_{+}(b))  R^{m_0}
k^{-\tau}.\nonumber
\end{align}
The first term in \eqref{eq:split2} is easily estimated using
\eqref{eq:quantum_part} and
\eqref{eq:quantum_part_iterated} with $m=m_0$.
The second term is estimated by combining
\eqref{eq:quantum_part} and
\eqref{eq:PsDO_part_iterated} (with $m=m_0$).
Finally, the third term is estimated, using the support
properties of $G_{-}$, by combining
\eqref{eq:disjoint_a} and
\eqref{eq:x_weights_iterated} (with $m=m_0$).

{\bf Proof of \eqref{eq:PsDO_part_iterated}}.
We shall use a set of functions $G$'s as above such that 
$\tilde G_{-}
\equiv 1$ on a neighbourhood of $\supp \tilde F_{-}$.
 We
consider \eqref{eq:split2} now with $B=\Op(F_{-}(a_0)\tilde{F}_{-}(b))$.
The first term 
is bounded
using \eqref{eq:PsDO_part}
 and \eqref{eq:quantum_part_iterated}.
For the second term we apply
\eqref{eq:PsDO_part} and \eqref{eq:PsDO_part_iterated}.
Furthermore, using the support properties of
$\tilde{G}_{-}(b)$, we can apply
\eqref{eq:disjoint_b}  and \eqref{eq:x_weights_iterated}  to
bound the third term.

{\bf Proof of \eqref{eq:disjoint_b_iterated}}.
In order to prove \eqref{eq:disjoint_b_iterated}
we choose a set of functions $G$'s as above such that 
$G_{-} \equiv 1$ on a neighbourhood of $\supp F^{2}_{-}$,
$\dist(\supp \tilde{G}_{-}, \supp \tilde{F}_{+}) > 0$ and 
$\dist(\supp \tilde{G}_{+}, \supp \tilde{F}_{-}) > 0$. Now we
write with $B_1=\Op(F^{1}_{-}(a_0)\tilde{F}_{-}(b))$ and $B_2=\Op(F^{2}_{-}(a_0)\tilde{F}_{+}(b))$
\begin{align}
\label{split_6}
k^tB_1 
R^{m_0+1}
B_2k^t 
=&
k^t B_1
R
\Op(G_{+}(a_0))
R^{m_0}
B_2k^t \\
&+
k^t B_1
R \Op(G_{-}(a_0)\tilde{G}_{-}(b))
R^{m_0}
B_2k^t \nonumber \\
&+
k^t B_1
R \Op(G_{-}(a_0)\tilde{G}_{+}(b))
R^{m_0}
B_2k^t. \nonumber
\end{align}
To bound the first term in \eqref{split_6}, we use the
support property of $G_{+}$, \eqref{eq:PsDO_part}
and \eqref{eq:disjoint_a_iterated}. The second term is
bounded using the separation of the supports of
$\tilde{G}_{-}$ and $\tilde{F}_{+}$,
\eqref{eq:PsDO_part} and \eqref{eq:disjoint_b_iterated}.
Finally for  the third term we combine
\eqref{eq:disjoint_b},  \eqref{eq:PsDO_part2_iterated} and the
fact that the supports of $\tilde{F}_{-}$ and $\tilde{G}_{+}$
are separated.

{\bf Proof of \eqref{eq:disjoint_a_iterated}}.
We finally consider \eqref{eq:disjoint_a_iterated} with
$m=m_0+1$. Here we choose $G$'s as before, but this time
satisfying the conditions $\dist(\supp G_{+}, \supp F_{-}) >
0$, $\dist(\supp G_{-}, \supp F_{+}) >0$ and \newline $\dist( \supp
\tilde{G}_{-}, \tilde{F}_{+}) > 0$. Introducing the
corresponding partition of unity, we have to bound the terms in
\eqref{split_6} with
$B_1=\Op(F_{+}(a_0))$ and $B_2=\Op(F_{-}(a_0)\tilde{F}_{+}(b))$.
Since $\dist(\supp G_{+}, \supp F_{-}) >0$, we can use
\eqref{eq:quantum_part} and
\eqref{eq:disjoint_a_iterated} to bound the first term.  For the second term we use \eqref{eq:quantum_part},
$\dist( \supp
\tilde{G}_{-}, \tilde{F}_{+}) > 0$ 
and \eqref{eq:disjoint_b_iterated}. For the third term we
apply the support condition on $G_{-}$,
\eqref{eq:disjoint_a}, and \eqref{eq:PsDO_part2_iterated}.

\section{Uniformity of  H\"{o}rmander metric}
\label{verification}

In this appendix we verify that the metric $g_E=\frac{dx^2}{\langle x
  \rangle^2} + \frac{d\xi^2}{f_E(x)^2}$ satisfies the estimates in the
  definition of a H\"{o}rmander metric {\it uniformly} in the
  parameter $E\in \left( 0,1\right]$, cf. Lemma \ref
{lem:hormander_metric}.

\begin{proof} First we prove that for some $C>0$ independent of $E,x$
  and $y$
\begin{align}
  \label{eq:slow_comparable0}
 \frac{f(x)}{f(y)} \leq C \left( 1+\frac{\langle y \rangle }{\langle x
    \rangle }\right)^{\mu/2}.
\end{align}
Suppose $\langle x \rangle^{-\mu} \leq E$. Then 
$$
\frac{f(x)}{f(y)}
\leq \sqrt{ \frac{\kappa_0^{-2}E + (1-\mu/2)^{-1}E}{\kappa_0^{-2}E}} =
\sqrt{\frac{\kappa_0^{-2} + (1-\mu/2)^{-1}}
{\kappa_0^{-2}}}.
$$
On the other hand if $\langle x \rangle^{-\mu} \geq
E$, then
$$
\frac{f(x)}{f(y)}
\leq \sqrt{ \frac{\kappa_0^{-2}\langle x \rangle^{-\mu} + (1-\mu/2)^{-1}
    \langle x \rangle^{-\mu}}{(1-\mu/2)^{-1} \langle y
    \rangle^{-\mu}}} = C \left( \frac{\langle y \rangle}{\langle x
    \rangle}\right)^{\mu/2}. 
$$ 

To prove the slow variation, let us assume that $1/C_1 < 1/4$. The inequality
$g_{(x,\xi)}((y,\eta)) \leq 1/C_1$ implies $|y|^2 \leq \langle x
\rangle^2/C_1$ and therefore
\begin{align}
  \label{eq:slow_comparable}
   \frac{2}{3}\leq \frac{\langle x \rangle}{\langle x+y \rangle} \leq 2.
\end{align}
Now using \eqref{eq:slow_comparable0} and \eqref{eq:slow_comparable},
\begin{align*}
  g_{(x,\xi)+(y,\eta)}(v) & = \frac{v_x^2}{\langle x+y\rangle^2} +
  \frac{v_{\xi}^2}{f(x+y)^2} \\
  &\leq \sup\Big\{ \left( \frac{\langle x \rangle}{\langle x+y
  \rangle} \right)^2, \left( \frac{f(x)}{f(x+y)} \right)^2 \Big\}
  g_{(x,\xi)}(v)\\
&\leq \sup\Big\{ 4, C\left( \frac{5}{2} \right)^{\mu} \Big\}
  g_{(x,\xi)}(v).
\end{align*}

The dual metric is given by 
$$
g^{\sigma} = f^2 dx^2 + \langle x \rangle^2 d\xi^2.
$$
Since clearly 
$$
\langle x \rangle^{-2} \leq \langle x \rangle^{-\mu} \leq f^2,
$$
we see that the uncertainty principle is satisfied.

Finally, we prove that the metric is temperate. Let us first 
consider the case $|x|\geq |y|$. Then
$$
\frac{g_{(x,\xi)}(v)}{g_{(y,\eta)}(v)} = 
\frac{ 
\frac{v_{x}^2}{\langle x \rangle^2} + \frac{v_{\xi}^2}{f(x)^2} }
{\frac{v_{x}^2}{\langle y \rangle^2} + \frac{v_{\xi}}{f(y)^2}} =
\frac{a^2v_x^2 + b^2 v_{\xi}^2}{c^2v_x^2 + d^2 v_{\xi}^2},
$$
where $c^2 \geq a^2$, $b^2 \geq d^2$. One now sees that the function
$s \mapsto \frac{a^2 + b^2 s}{c^2+d^2 s}$ is increasing on
$[0,\infty)$. Therefore, using also \eqref{eq:slow_comparable0}, we infer that for all $v \neq 0$
$$
\frac{g_{(x,\xi)}(v)}{g_{(y,\eta)}(v)} \leq \frac{b^2}{d^2} =
  \frac{f(y)^2}{f(x)^2}\leq C^2\left( 1+\frac{\langle x \rangle }{\langle y
    \rangle }\right)^{\mu}\leq C^2\left( 1+\frac{\langle x \rangle }{\langle y
    \rangle }\right)^2.
$$
A similar argument shows that in the case $|x| < |y|$
$$
\frac{g_{(x,\xi)}(v)}{g_{(y,\eta)}(v)} \leq \frac{a^2}{c^2} =
  \frac{\langle y \rangle^2}{\langle x \rangle^2}.
$$
We need to find $\tilde C, N>0$ such that
\begin{align}
\label{eq:tempered_final1}
 \max\Big\{ \frac{\langle x \rangle^2}{\langle y \rangle^2},
\frac{\langle y \rangle^2}{\langle x \rangle^2}\Big\}\leq
\tilde C \Big(1 + g_{(x,\xi)}^{\sigma}((y,\eta)-(x,\xi))\Big)^N. 
\end{align} 
Clearly 
$$
g_{(x,\xi)}^{\sigma}((y,\eta)-(x,\xi)) \geq f(x)^2 |y-x|^2 \geq c
\langle x\rangle^{-\mu} (|x|-|y|)^2.
$$
If $|x| \geq 2 |y|$, then 
$$
\langle x\rangle^{-\mu} (|x|-|y|)^2 \geq 4^{-1} \langle x\rangle^{-\mu} |x|^{2}.
$$
Using the trivial bound $\frac{\langle x \rangle^2}{\langle y \rangle^2} \leq \langle x
\rangle^2$ we conclude that in this case any $N \geq 2/(2-\mu)$
suffices.
If $|x| \leq 2^{-1} |y|$, then
$$
\langle x\rangle^{-\mu} (|x|-|y|)^2 \geq 4^{-1} \langle y\rangle^{-\mu} |y|^{2},
$$ and again $N \geq 2/(2-\mu)$
suffices.
  If, on the other hand,
$2^{-1} |y|<|x| < 2 |y|$, then $\max\{ \frac{\langle x \rangle^2}{\langle y \rangle^2},\frac{\langle y \rangle^2}{\langle x \rangle^2}\}\leq 4$. 

We have
proved \eqref{eq:tempered_final1} and hence that the metric is temperate uniformly in $E$. That
finishes the proof of the first part of Lemma
\ref{lem:hormander_metric}.

 The second part of Lemma
\ref{lem:hormander_metric} for $m=\langle x\rangle$ or $m=f_E$ follows readily from 
\eqref{eq:slow_comparable0},~\eqref{eq:slow_comparable} and
\eqref{eq:tempered_final1}.
The statement for $m=\langle \xi \rangle$ or $m=\langle \frac{\xi}{f_E}
\rangle$ may be verified along the same line; we omit the proof. The statement
for products of powers is a general property for uniformly temperate
weight functions.

\end{proof}
 
   \bibliographystyle{amsalpha}

\end{document}